%% file: HackMotors.tex
\documentclass[sigconf]{acmart}

\setcopyright{none} 
\acmYear{2022}

\usepackage[abbreviations = true]{siunitx}
\usepackage{graphicx}
\graphicspath{{./Figures/block/}{./Figures/plots}}
\usepackage{import}
\usepackage{calc}

\usepackage{balance}
\usepackage[all]{nowidow}

\begin{document}
\title{Detection of Electromagnetic Signal Injection Attacks on Actuator Systems}
\author{Youqian Zhang}
\email{youqian.zhang@cs.ox.ac.uk}
\affiliation{%
  \institution{University of Oxford}
}
\author{Kasper Rasmussen}
\email{kasper.rasmussen@cs.ox.ac.uk}
\affiliation{%
  \institution{University of Oxford}
}

\begin{abstract}
An actuator is a device that converts electricity into another form of energy, typically physical movement. They are absolutely essential for any system that needs to impact or modify the physical world, and are used in millions of systems of all sizes, all over the world, from cars and spacecraft to factory control systems and critical infrastructure.
An actuator is a ``dumb device'' that is entirely controlled by the surrounding electronics, e.g., a microcontroller, and thus cannot authenticate its control signals or do any other form of processing. The problem we look at in this paper is how the wires that connect an actuator to its control electronics can act like antennas, picking up electromagnetic signals from the environment. This makes it possible for a remote attacker to wirelessly inject signals (energy) into these wires to bypass the controller and directly control the actuator.

To detect such attacks, we propose a novel detection method that allows the microcontroller to monitor the control signal and detect attacks as a deviation from the intended value. We have managed to do this without requiring the microcontroller to sample the signal at a high rate or run any signal processing. That makes our defense mechanism practical and easy to integrate into existing systems. Our method is general and applies to any type of actuator (provided a few basic assumptions are met), and can deal with adversaries with arbitrarily high transmission power. We implement our detection method on two different practical systems to show its generality, effectiveness, and robustness.
\end{abstract}



\keywords{electromagnetic signal injection; embedded system security; actuator} 

\maketitle
\section{Introduction}
\label{sec:introduction}

Actuator systems are embedded in our daily lives to such an extent that it is hard to find an example of an electronic system that does not have actuators in some form. Anything from household devices like smart locks or robotic vacuum cleaners, to transportation to production to defense. Actuators are everywhere.

These devices interact with the physical world by converting an electric signal into some other form of energy, typically movement, but a heater or a light source can also be considered an actuator. It is well known that the wires used to feed electrical signals to such devices can act as antennas, unintentionally picking up electromagnetic interference (EMI)~\cite{kune2013ghost,leone1999coupling,goedbloed1992electromagnetic} from the environment, or indeed from an attacker. This inherent vulnerability allows an attacker to \emph{wirelessly} inject attacking signals into the wires, disturbing or changing the original signals.

Since an actuator is simply an energy transducer, it cannot authenticate its input signals and will respond to whatever it receives, in the worst case resulting in the adversary being able to fully control the state of the actuator. It is easy to see how such an attack can be used, e.g., to rotate the motor in the smart lock to unlock a door; or force closed a fuel injection valve in a car to stop the car's engine. When the target system is complex and important, these attacks can be incredibly powerful and dangerous. For example imagine the potential harm if an adversary could control critical industrial applications (e.g., robotic arms) or medical devices (e.g., pacemakers), or say, move the control surfaces of an airplane without pilot input.

Even though such attacks are complicated to perform in practice, and as a result are still rare, we need to find effective detection and mitigation strategies to deal with them before they become common. 
In the last few years, there has been work on detecting such attacks on sensors~\cite{sp20Zhang,shoukry2015pycra,kohler2021signal,ruotsalainen2021watermarking, fang2022detection}. 
In this paper, we focus on detecting attacks on actuators, which is quite a bit harder. 
The reason is that when a sensor is attacked, the receiving device is a microcontroller that has the ability to run filters and algorithms, or use redundant measurements for added security. For actuators that is not as easy. When actuators are attacked the receiving device is the actuator itself, and since actuators are ``dumb devices'' (it might just be a coil of wire, like in a motor), they do not have the ability to ignore malicious signals, even if such signals deviate from some usual pattern.
 
In this paper we provide a novel detection method that uses common and inexpensive electrical components, making it possible to apply our method at scale. The basic idea is to compare the signal to be protected with a reference, in order to identify when any external interference is present. However, this is not as easy as it sounds. First of all, an adversarial signal will affect any reference signals as well, and there are challenges with the sampling rate, bandwidth limits, and signal processing efforts that can make a trivial scheme unusable in practice. Our detection method solves all those problems and we are able to provide strong detection guarantees for (almost) any actuator system.

Our contributions of this work are summarized as follows:
\begin{itemize}
\item We create a universal and flexible system model that fits most (if not all) actuator systems. It allows us to capture the specific needs of any system by tuning parameters of the model (Section~\ref{sec:actuator_system}).
\item We propose a general and lightweight detection method that uses differential amplifiers to detect electromagnetic signal injection attacks, and we show that it can provide the actuator system with a strong security guarantee (Section~\ref{sec:attack_detection} and Section~\ref{sec:protection_outofthe_operational_band}).
\item We implement the proposed detection method on a speaker system and a motor control system, and we demonstrate the generality, feasibility, and robustness of our detection method (Section~\ref{sec:implementation}). 
\end{itemize}

In the remaining parts of this work, we present a background on the electromagnetic signal injection attacks in Section~\ref{sec:background}.
Furthermore, additional important issues are discussed in Section~\ref{sec:discussion}, and related work is summarized in Section~\ref{sec:related_work_existing_defenses}.
Finally, a conclusion of this work is drawn in Section~\ref{sec:conclusion}.

\section{Background}
\label{sec:background}

Before introducing our detection method, we first provide a brief background on electromagnetic signal injection attacks. 
We first explain how electromagnetic waves are injected into a victim system, and next, we explain how the injected signals influence the actuator, as well as presenting successful attacks in previous studies.

\subsection{Electromagnetic Signal Injection}
Electromagnetic fields can affect a metal conductor by inducing voltage changes into it, and this has been thoroughly studied in the area of ``Electromagnetism''.
Besides antennas for wireless communications, the metal conductor also exists in devices in the form of wires (or traces) that connect electronic components.
These wires can also act as antennas to capture environmental electromagnetic waves.
Many researchers exploited such ``antenna-like'' behaviors of the wires to radiate electromagnetic waves and wirelessly inject them into the circuits~\cite{kasmi2015iemi,kune2013ghost,
Kasper2009PACf,selvaraj2018electromagnetic,
Markettos2009Tfia, osuka2018information,giechaskiel2019framework, shin2016sampling, tu2019trick, sp20Zhang,wang2022ghosttouch, dayanikli2020senact, dayanikli2021electromagnetic,ware2017effects,selvaraj2018intentional}.

Many factors affect the injection process, but the attack power and the attack frequency are the basic ones that an attacker tunes, as they determine the effectiveness and efficiency of the injection~\cite{yan2020sok}.
To cause effective impacts onto the circuits, the injected voltage needs to be strong enough.
Since the injected voltage is proportional to the attack power~\cite{friis1946note}, the more powerful the attacking signal is, the higher the injected voltage will be, and it is more likely the attack is effective. 
In addition, in order to maximize the injected voltage, the attack frequency must be the resonant frequency of the wire; at other frequencies, it will cost more attack power to achieve the same amount of injected voltage~\cite{kune2013ghost}. 
By properly tuning the attack power and the attack frequency, the attacker can inject arbitrary signals into the wires.

\subsection{Circuit Response to Injected Signal}
After the injection, a successful attack depends on how the circuits respond to injected signals.
On the one hand, the injected signal can be within the frequency band in which the circuits are designed to operate, namely the operational band (in-band).
Since the malicious voltage changes are within the operational band, the circuits respond to them directly. This will subsequently influence a signal that drives the actuator, further impacting the actuator responses.
On the other hand, the injected signal can also be out of the operational band (out-of-band). 
However, in order to affect the circuits in an effective and predictable way, it is essential to cause voltage changes within the operational band.
A well-studied method of transferring the out-of-band changes to the operational band is exploiting the nonlinear properties of electronic components: the attacker first exerts an in-band malicious signal onto an out-of-band radio-frequency (RF) carrier to form the attacking signal; next, after the signal injection, the malicious signal is extracted from the attacking signal due to nonlinearities of electronic components such as amplifiers~\cite{kune2013ghost, kasmi2015iemi, tu2019trick}, electro-static discharge (ESD) circuits~\cite{selvaraj2018electromagnetic, dayanikli2020senact}, and analog-to-digital converters (ADCs)~\cite{giechaskiel2019framework, giechaskiel2019sok};
as a result, the in-band malicious signal appears in the operational band, further affecting the actuator.

Here are some examples of manipulating actuators. Selvaraj et al.~\cite{selvaraj2018electromagnetic} demonstrated how to inject fine-tuned attacking signals into the target wire to precisely manipulate servo angles.
They exploited the nonlinearities of the electro-static discharge (ESD) circuits to toggle the voltage level of the signal so that they can precisely control the signal pulse width that determines the servo angle.
Danyanikli et al.~\cite{dayanikli2020senact} demonstrated similar attacks: they could remotely manipulate the signal that controls switches in an AC-to-DC power converter, which regulates power delivery to electric vehicles.
The attack can forcibly toggle the on/off state of the switch. 
An irreparable result of the attack is causing a short circuit to burn the converter.
In this work, we will demonstrate that an arbitrary audio signal can be injected into a speaker system, in which the nonlinearities of the audio amplifier is exploited (see details in Section~\ref{sec:speaker}); furthermore, we will show that a motor control system can be precisely controlled, in which the imperfection of transistors~\cite{bona2010new, bona2009eaa, FioriFranco2014SoSP} are exploited (see details in Section~\ref{sec:motor}).
All of these show that it is not difficult to use electromagnetic signal injection attacks to manipulate the actuator behaviors precisely.

\begin{figure}[t]
  \def\svgscale{0.55}
  \centering
  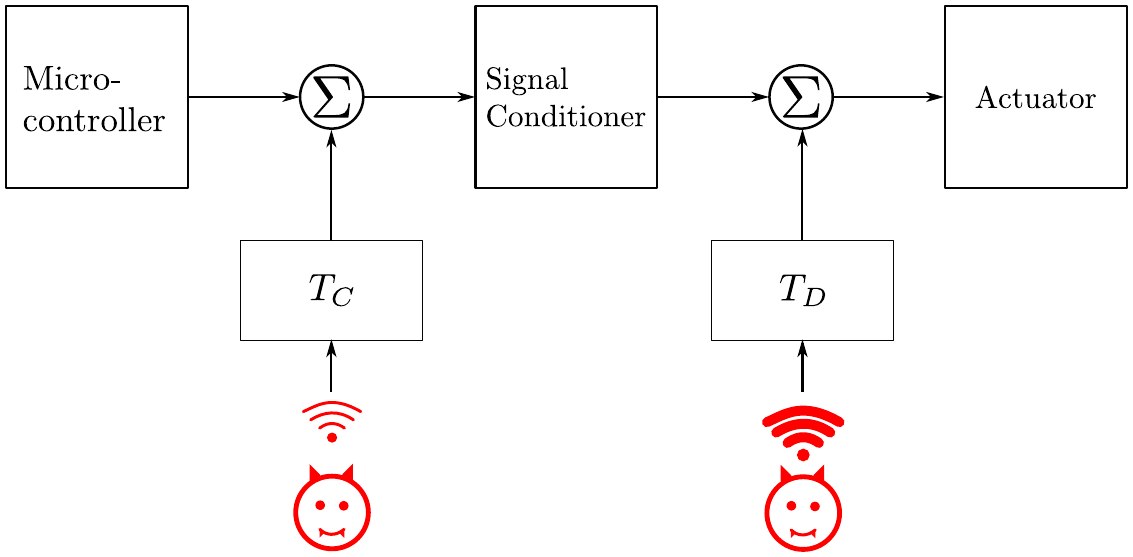
  \caption{The actuator system consists of a microcontroller, a signal conditioner, and an actuator. The transfer functions $T_{C}$ and $T_{D}$ explain the control signal wire and the drive signal wire capturing the attacking signals, respectively.}
  \label{fig:system_model}
\end{figure}

\section{System Model and Adversary Model}
\label{sec:actuator_system}

In this section, we introduce a general and flexible system model that fits most actuator systems. 
This allows us to capture the needs of any specific system by tuning the model parameters.
We also present a comprehensive attacker model that, together with the system model, forms a flexible tool to describe signal injection attacks and defense mechanisms on actuator systems.

\subsection{System Model}
\label{sec:system_model}

We call a system that controls an actuator an ``actuator system''.
In an actuator system, a microcontroller is the device used to regulate, command, and manage the behaviors of the actuator.
Between the microcontroller and the actuator, there are circuits transforming the microcontroller output signal into a suitable signal to drive the actuator.
For instance, such circuits may be for signal amplifications or waveform conversions.
To capture all the characteristics of the circuits, we define a new device, called the signal conditioner.
How this device works will differ from circuit to circuit, but we treat it as a black box.
Therefore, our system model consists of three devices: a microcontroller, a signal conditioner, and an actuator; a block diagram of the system model is presented in Figure~\ref{fig:system_model}.

In the system model, wires are used to connect these devices: as shown in Figure~\ref{fig:system_model}, one wire is used to transmit the microcontroller output signal, which we call the \emph{control~signal}, to the signal conditioner; the other transmits the signal conditioner output signal, which we call the \emph{drive~signal}, to the actuator.
Note that, in practice, the control signal wire often carries comparatively little power compared to the drive signal wire, as the voltage and the current of the microcontroller outputs are constrained to a few volts and milliamperes, respectively. Whereas the drive signal wire can carry high-power signals because some actuators consume significant power while working.

The operational frequency of the control signal and drive signal can vary significantly from system to system.
Still, it is generally possible to define a normal operating range to which signals are confined in normal operation. 
This is important because while low/high pass filters can filter out adversarial injections at extreme frequencies, it is more difficult to filter out attacks in the operational range without affecting the valid control/drive signal.
Our solution assumes that such an operational range can be defined and we call the upper limit of this range $f_{max}$. 
Note that we make no assumptions about what the value of this limit is, only that it exists.
In Section~\ref{sec:protection_outofthe_operational_band}, we discuss ways of extending this range way beyond the design limits of the electrical components.

\subsection{Adversary Model}
\label{sec:adversary_model}

The attacker's goal is to affect the actuator by electromagnetic interference, i.e., inject an attacking signal into the system.
The attacker can inject attacking signals into the actuator system remotely but cannot physically access or modify the actuator system.
We grant the attacker full knowledge of the actuator system; specifically, the attacker can predict the waveform and timing (phase) of signals in the actuator system.
The attacker can also craft any (physically possible) signal she wants. 

In practice, signal injection can be rather complicated, especially from far away, but we deliberately grant the adversary extremely strong power to make sure our detection method works in every case. 
We manage this complexity using a transfer function that encapsulates any changes to the attacking signal caused by the injection process, e.g., frequency selectivity, attack distance, attenuation, spreading and convolution, etc.,\ as shown in Figure~\ref{fig:system_model}.
We do not limit the power available for the adversary. 
However, we do assume that there exists a lower limit, below which any injected signal no longer has a meaningful effect on the target system. 
We call this lower limit $P_{min}$. 
This power limit is set by the system designer to make sure that any injected signal above this limit is detected. 
It can be set arbitrarily low, but in order to successfully attack the system, the attacker must inject a signal with power higher than $P_{min}$.

The reason why we grant the attacker ideally strong abilities is that if such an attacker cannot avoid being detected by our proposed detection method, it is impossible for any other attackers who are no better than this ideally strong attacker to bypass the detection method.

\subsection{Two Injection Points}
\label{sec:two_injection_points}

In a particular physical system, there could exist multiple injection places through which attacking signals enter the system.
Therefore, many electronic components will also be affected by the injections.
However, only when these injections lead to effects on signals that directly determine responses of the system will the system be successfully manipulated by the attacker, and this has also been considered and shown in previous studies~\cite{tu2019trick, esteves2018remote}.
Therefore, regardless of where the signal injection happens in the system, even if currents are induced in many places at once, it is possible to find an input signal that, when applied to one of the two wires in our model, produces the same effect.
This means that without loss of generality, we can model any signal injection as if the attacking signal was injected into the control or drive signal wire through an appropriate transfer function. 
In practice, signal injection does in fact almost exclusively happen via these wires, because these are the most efficient ``antennas'' in the system, and thus where most of the energy is transferred.

There is an important difference between these two injection points. As mentioned previously, the power of the control signal is comparatively weak, so the adversary can more easily overshadow any valid signal on the control signal wire, and it will generally take less power to make changes that affect the actuator through this injection point. We define such an injection as a \emph{control~signal~injection}. 

The second injection point is the drive signal wire. This wire will generally carry signals with higher power and more specialized waveforms. For some actuators, e.g., brushless electric motors, the drive signals are not only high-powered but also somewhat complex, and the timing between the different phases of the signal is very important to the operation of the motor. This means that injection into this wire is more difficult and requires much more power from the adversary. We define such an injection as a \emph{drive~signal~injection}.

Despite this difference in attacker capabilities between the two injection points, our detection system, described in the following sections, works for both the control signal wire and the drive signal wire.

\section{Attack Detection}
\label{sec:attack_detection}

\begin{figure}[t]
  \def\svgscale{0.72}
  \centering
  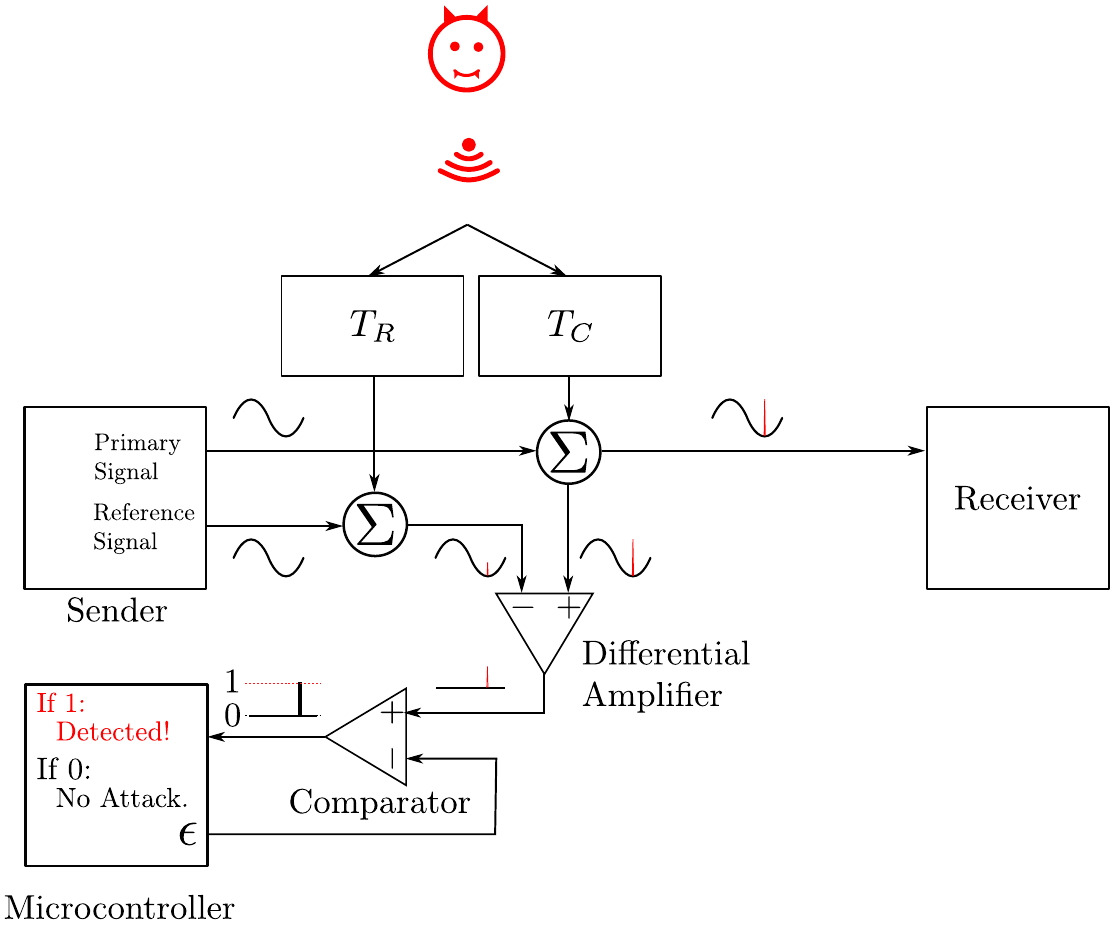
  \caption{A differential amplifier compares the primary signal that is transmitted from the sender to the receiver with the reference signal. A comparator circuit further compares the differential amplifier output with a threshold $\epsilon$, and the microcontroller determines whether attacks happen according to the binary results of the comparator.}
  \label{fig:add_comparator}
\end{figure}

As mentioned previously, our detection approach works for both the control signal wire and the drive signal wire. Rather than choose one of the two injection points for our description, we instead treat each wire as having a ``sender'' and a ``receiver''. 
The sender device is thus either the microcontroller or the signal conditioner, with the corresponding receiver being the signal conditioner or the actuator, respectively. We discuss any minor differences between the injection points in Section~\ref{sec:differences_injection_points}.

In order to detect a signal injected into the wire, we make the sender generate two identical signals, one primary signal (sent to the receiver) and one reference signal (used for detection). This idea is illustrated in Figure~\ref{fig:add_comparator}. A differential amplifier is then used to amplify any differences between the primary and reference signals. In the absence of an attack, these two signals should be identical and thus produce no output from the amplifier. However, if the primary wire is affected by an external signal, the difference will be amplified and can be detected using a comparator and a microcontroller.  A very important requirement is that the reference wire is sufficiently different from the primary wire to make sure that the adversary cannot modify both in the same way. This can be easily accomplished by simply making the wires different lengths~\cite{balanis2016antenna} in order to make them sensitive to different frequencies, but more significant difference can be achieved, e.g., with additional Radio Frequency (RF) shielding materials on the reference wire. 

When no attack signal is present, the two input signals of the differential amplifier (the primary and the reference) are the same, and the differential amplifier output is zero. In reality, there will be a non-zero amount of noise, but the output is essentially zero. When an attack happens, the primary wire and the reference wire both pick up the attacking signal. But, because the two wires cannot be modified in the same way, captured by the two different transfer functions $T_C$ and $T_R$ shown in Figure~\ref{fig:add_comparator}, the two inputs to the differential amplifier will be different. This results in a non-zero signal on the output of the differential amplifier, and allows the microcontroller to detect the attack.

It is essential to emphasize that the differential amplifier is used in a novel way that is different from how it is commonly used in analog electronics, in which the differential amplifier is used to reduce equal interference (common-mode interference) onto its two inputs~\cite{razavi2005design}. However, in our detection method, the two inputs are deliberately crafted such that the differential amplifier captures the attack interference, rather than mitigate it.

\subsection{Modeling Differential Amplifier Output}
\label{sec:comparator_response}

The differential amplifier amplifies the difference of its input signals. We model this as the difference between the primary and the reference $\delta(t)$, plus additive white Gaussian noise $n(t)$, amplified by a constant gain $G$. To simplify the notation, we omit $t$ hereafter. The output of the differential amplifier becomes:
\[ o = G(\delta + n) \]

Given an attacking signal $s$, the signal that is injected in the primary wire is $T_{C}(s)$, and the signal that is injected in the reference wire is $T_{R}(s)$.
In order to obtain a simple relationship between these two injected signals, we make the simplifying assumption that $T_{R}$ can be expressed as being $K$ times weaker than $T_{C}$. Therefore we can write:
\[ \delta = T_{C}(s) - T_{R}(s) = T_{C}(s) - \frac{1}{K}T_{C}(s) = \frac{K-1}{K}T_{C}(s) \]

Thus, the output of the differential amplifier becomes
\begin{align}
\label{eq:o_tc}
o = G\left(\frac{K-1}{K}T_{C}(s) + n\right)
\end{align}

Finally we take advantage of the fact that the power that is absorbed by the receiving antenna (the primary wire) is proportional to the attack power $P$~\cite{friis1946note}, to arrive at the final model for our detection system:
\begin{align}
  o = G\left(\frac{K-1}{K} P  + n\right)
  \label{eq:o_p}
\end{align}

This equation gives us the output of the differential amplifier as a function of the main parameters of our detection system, namely the noise $n$, the gain of the differential amplifier $G$, the ``difference'' of the primary and reference wires $K$, and the power of the adversarial signal $P$.

\subsection{Detection Rule and Choice of Parameters}
\label{sec:detection_procedures}

According to Equation~\eqref{eq:o_p}, when no attack signal is present, i.e., the attacker's power is 0, we have ${o=Gn}$. To make sure that small amounts of noise do not cause false positives, we define a threshold $\epsilon$ that the output of the amplifier must exceed in order to be detected as an attack. The actual detection is done by a comparator whose output is high when $o\ge\epsilon$ and low otherwise. This allows the output of the comparator to be fed into an interrupt pin of the microcontroller, as shown in Figure~\ref{fig:add_comparator}, and ensures that even attack signals with a very short duration can be detected efficiently without requiring the microcontroller to sample at a high rate.

\begin{figure}[tb]
  \centering
  \includegraphics[width=.5\linewidth]{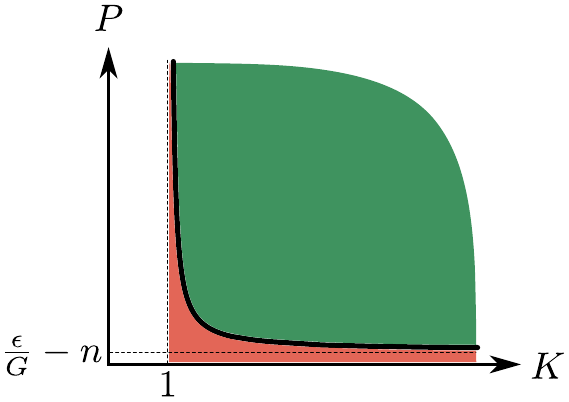}
  \caption{The minimum detectable attack power is expressed as a function of $K$. The detection method can detect attacks on and above the curve (green), but it cannot detect attacks below the curve (red). By decreasing (increasing) $\epsilon$ or increasing (decreasing) $G$, the horizontal asymptote can be moved down (up).}
  \label{fig:injected_vs_k}
\end{figure}

Since the attack is detected if:
\[ G\left(\frac{K-1}{K} P  + n\right) \ge \epsilon \]
we can rearrange this to see that the detection method can detect any attack with power that fulfills the following requirement:
\begin{align}
  P \geq \left(\frac{\epsilon}{G}-n\right) \cdot \frac{K}{K-1}
  \label{eqn:min_power}
\end{align}

From Inequality~\eqref{eqn:min_power} we see that the minimum detectable power can be made arbitrarily small with appropriate choices of $K$, $G$, and $\epsilon$. In the following, we describe the procedures for choosing these values.

The choice of $K$ is relatively simple: bigger is better. A large $K$ means that the difference between the two transfer functions $T_C$ and $T_R$, which govern how an attacking signal affects the primary and references wires, is as big as possible. To get a sense of how the choice of $K$ affects the detection performance, we plot the attacker's power $P$ as a function of $K$ in Figure~\ref{fig:injected_vs_k}. The detection method detects attacks on or above the curve (green region), while the attacks below the curve (red region) are not detected. We see that $K$ does not have to be very high for the detection method to be effective, but it does have to be above 1, i.e., the primary and reference wires do have to differ.

The amplification of the differential amplifier $G$ is dictated by the choice of amplifier used. Different amplifiers have different maximum gains, and typical values range from 100 to 300. Generally, $G$ should be chosen as high as possible, although in noisy environments, it may be beneficial to reduce the amplification to reduce the sensitivity to noise. 

As for choosing the detection threshold $\epsilon$, it needs to be chosen such that environmental noise does not cause a detection event.
Therefore, $\epsilon$ is chosen just high enough to make sure that false positives from noise are kept to a minimum; we show an example of this in our implementation in Section~\ref{sec:implementation}.
Moreover, because noise environments are often complicated and change significantly over time, we emphasize that $\epsilon$ does not have to be constant. It can for example be adaptively adjusted to accommodate lower levels of ambient electromagnetic noise during the night. We further discuss this in Section~\ref{sec:adaptive_threshold}.

\subsection{Security Analysis}
\label{sec:security_analysis}

Recall from the adversary model that the goal of the attacker is to affect the actuator. In order to achieve this goal, the attacker must inject a signal with power of at least $P_{min}$. We prove that such attacks are always detected by our detection method as follows.

Substituting $P_{min}$ into Inequality~\eqref{eqn:min_power} we see that if the following inequality holds the attack is detected:
\[ P_{min} \ge \left(\frac{\epsilon}{G}-n\right) \cdot \frac{K}{K-1}\]

To show that it is always possible to find values of $K$, $\epsilon$, and $G$ to make this inequality true, for any value of $n$ we first make the observation that $K$ can be made arbitrarily high independent of noise. Since $K/(K-1)$ approaches 1 for high enough values of $K$, we can pick a high value and reduce the above inequality to
\[ G(P_{min} +n) \ge \epsilon\]

In addition, as mentioned in the previous subsection, it is a functional requirement that the detection threshold must not be triggered by the noise alone, i.e., the following must hold:
\[ Gn < \epsilon\]

Both of the two inequalities above must be true in order to have a functional detection system. That gives the following constraint:
\begin{align*}
  G(P_{min} +n) &\ge \epsilon > Gn \\
  P_{min} +n &> n\\
  P_{min} &> 0
\end{align*}
Thus for any value of $P_{min} > 0$, it is possible to find values of $K$, $G$, and $\epsilon$ that allows the detection system to detect any adversarial signal with power above $P_{min}$ and at the same time do not trigger false positives from noise.

\subsection{Differences Between Injection Points}
\label{sec:differences_injection_points}

Recall that one injection point is the control signal wire, and the other is the drive signal wire.
The first difference is between the differential amplifiers at these two injection points.
Recall that the control signal has a low voltage level, and as such, it is sufficient for the differential amplifier to have an input voltage range of several volts.
However, the drive signal's voltage can go up to hundreds of volts, e.g., $\SI{380}{\volt}$ industrial motors. 
Thus, a differential amplifier with a large enough voltage input range is needed such that the tapped signal will not cause any damage to the differential amplifier.
It is not hard to find such a differential amplifier in the market.
Note that since the differential amplifier has a much higher impedance than the actuator, the tapping only draws a tiny portion of the control/drive signal, causing negligible impacts on the signal conditioner/actuator.

Another difference at these two injection points is that the drive signal can be much more complex than the control signal, and thus, it may be more complicated while deploying our approach for the drive signal.
In the previously mentioned example of a brushless electric motor, the microcontroller produces one signal for controlling, while the signal conditioner needs to convert this solitary control signal into three different signals to drive the motor.
In general, it is essential to deploy our approach to each signal to guarantee the security, which means one for the control signal and three for the drive signals.
However, in many cases where the physical properties of the multiple wires are the same or very similar and they are put close to each other, it is tricky that protecting one wire is sufficiently enough, and doing so can significantly reduce the complexity of deploying our approach.
This is because the attacker cannot selectively choose a wire to affect in these cases, and in other words, all of these identical or similar wires will be impacted by the attack.
In the example of the brushless DC motor, its three drive signal wires are almost identical, and they are put very close to each other.
Therefore, protecting any one of the wires with our approach is equivalent to protecting all three wires.

\subsection{Attacks on Detection Circuit}
\label{sec:attacks_on_detection_circuit}
Our defense mechanism has added circuitry to the system that could
itself be the target of an injection attack. In this section, we
demonstrate that this circuitry cannot be exploited by the attacker to
achieve the injection.

First, we note that there is no path from our detection circuit to the
actuator, so the only malicious action we have to consider is whether
an adversary could inject a signal that would be hidden from detection
because of interference in the detection circuit itself.

To analyze this, we define a new transfer function $T$ for the main
wire in the detection circuit. The resulting signal that is injected into the
detection wire is then $T(s)$ when the adversary sends $s$.
Note that there may also be multiple injection points as discussed in Section~\ref{sec:two_injection_points}, but they can be modeled to the main wire, as $o$ directly determines whether an attack happens or not. 
Therefore, the injected signal is superimposed onto $o$, described in
Equation~\ref{eq:o_tc}, making the modified differential amplifier
output~$o'$:
\begin{align*}
o' &= T(s) + G\left(\frac{K-1}{K}T_{C}(s) + n\right) \\
&= \frac{G(K-1)}{K}\left( \frac{K}{G(K-1)}T(s) + T_{C}(s)\right) + Gn
\end{align*}

If the attacker wants to avoid detection~$o'$ must be zero (technically just less than~$\epsilon$, but basically zero). That means that the value in the parentheses must be zero, which in turn requires the following equation holds:
\begin{align}
\label{eq:ts_tcs}
{\frac{K}{G(K-1)}T(s) = -T_{C}(s)}
\end{align}

The negative sign in Equation~\ref{eq:ts_tcs} implies that $T(s)$ and $T_{C}(s)$ must be 180 degrees out of phase, and this requires that the physical distance between the two corresponding wires is exactly half of the wavelength of the attacking signal $s$~\cite{balanis2016antenna}. This is already a good argument for why an attacker cannot inject a signal that affects the actuator, and simultaneously cancel it out in the detection system, since the frequency would have to be in the 10-100s of GHz to get a half wavelength short enough. Such a high-frequency signal is way above what affects most actuators.

However just to make the point extra clear, let's assume that the attacker could in fact send a signal with a high enough frequency to make this work. Even then, the constant $\frac{K}{G(K - 1)}$ in Equation~\ref{eq:ts_tcs} means that the signal injected into the detection system, $T(s)$, must be 100s of times stronger than $T_{C}(s)$, the signal injected into the actuator control signal itself.
However, given that the two wires are so close to each other, and that the smaller of the two needs more power injected into it, it is impossible to achieve such a $T(s)$ in practice. As a result Equation~\ref{eq:ts_tcs} can never hold in practice.

For those two reasons (phase difference and relative power), no adversarial
signal $s$ can ever prevent its own detection due to interference with
the detection circuitry itself.
\section{Extended Maximum Detectable Frequency}
\label{sec:protection_outofthe_operational_band}

\begin{figure}[t]
	\def\svgscale{0.65}
	\centering
    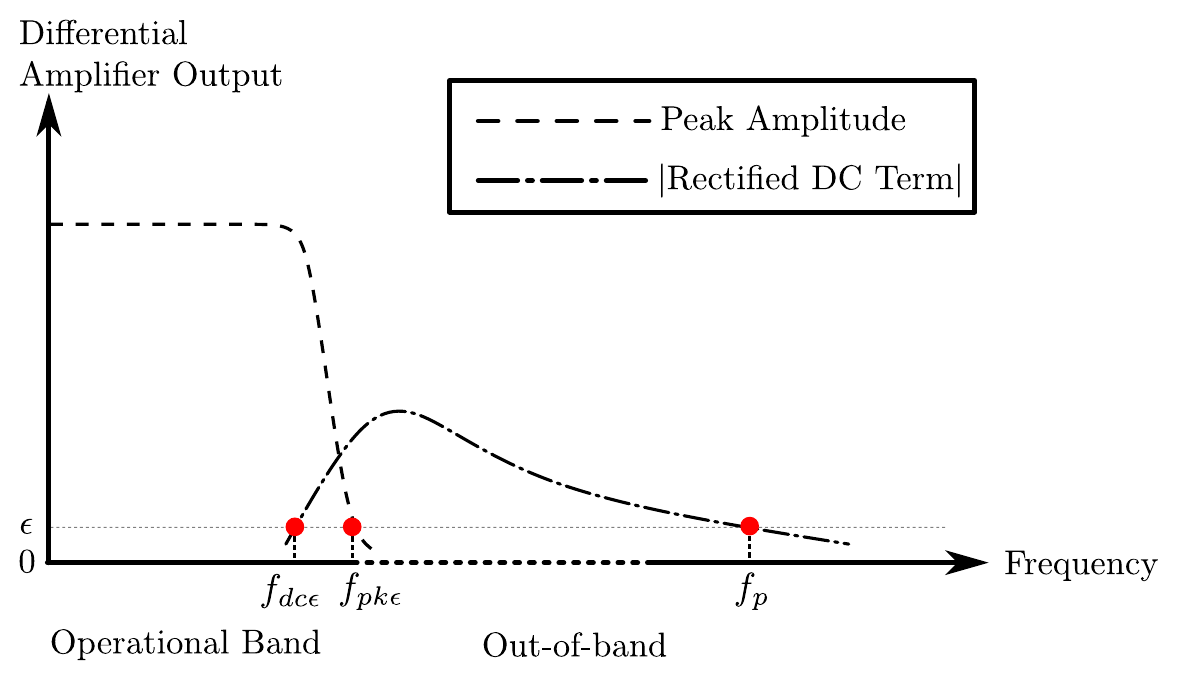
    \caption{With constant attack power, the peak amplitude (dashed line) and the rectified DC term (dash-dotted line) of the differential amplifier output signal change along with the frequency. The maximum detectable frequency is extended to $f_{p}$.}
    \label{fig:comparator_vs_attacking_signal}
\end{figure}

Our detection method relies on a differential amplifier to help detect injected signals. 
Like all electronic components, a differential amplifier is designed to work within a particular frequency range. 
When choosing parameters for the detection system, a suitable differential amplifier should be used, which covers the entire range where adversarial signals are likely to be able to affect the actuator. 
However, on rare occasions, e.g., for very high frequency applications or if cost is a significant concern, it might be difficult to get a differential amplifier that fully covers the desired frequency range. 
For such cases, we have come up with a method to extend the maximum detectable frequency $f_{max}$ beyond the normal upper bound of the differential amplifier.

Many previous studies~\cite{wu2018characterization, richardson1979modeling,forcier1979microwave,larson1979modified} have shown that a differential amplifier will still respond beyond its normal operational band, although the response is entirely different from the normal amplification within its design parameters.
As the frequency increases beyond $f_{max}$, the peak amplitude of the differential amplifier output starts to decline, as the gain plummets to almost zero~\cite{kitchin2006designer, mancini2003op}.
This is shown in Figure~\ref{fig:comparator_vs_attacking_signal} where the dashed curve depicts the change of the peak amplitude.

Although the peak amplitude decreases to nothing, the output will gain a DC offset with respect to the normal ground state, shown in Figure~\ref{fig:comparator_vs_attacking_signal} as the Rectified DC Term. 
This happens as the differential amplifier rectifies the high frequency signals~\cite{devices2009rfi,wu2018characterization,FioriFranco2014SoSP}. 
The phenomenon is also known as radio-frequency (RF) rectification, and it is attributed to the nonlinear voltage-current characteristic of transistors that make up the differential amplifier~\cite{devices2009rfi}. 
Further increasing the frequency will eventually decrease the rectified DC term, which will ultimately become negligible when the frequency is high enough~\cite{larson1979modified, richardson1979modeling,forcier1979microwave}. 
While this effect does eventually disappear, it allows us to extend the detection by hundreds or thousands of times higher than the upper bound of the operational band.

It is important to note that this phenomenon is not limited to a specific differential amplifier, but is true for many different designs, which has been experimentally verified in the literature~\cite{wu2018characterization, oapm2013specification}.

For our detection system to provide firm guarantees, it is essential to ensure no gap in the protected frequency band.
Therefore we have to ensure that the DC offset rises enough to be detected before the normal peak amplitude of the differential amplifier goes to zero. 
In Figure~\ref{fig:comparator_vs_attacking_signal}, the frequency at which the magnitude of the rectified DC term exceeds $\epsilon$ is denoted as $f_{dc\epsilon}$, and the frequency at which the peak amplitude falls below $\epsilon$ is $f_{pk\epsilon}$. We show in Section~\ref{sec:implementation} that we can easily achieve $f_{dc\epsilon} < f_{pk\epsilon}$ in practice.

\section{Implementation}
\label{sec:implementation}

We implement our detection method on two practical and distinct actuator systems: a speaker system (in Section~\ref{sec:speaker}) and a motor control system (in Section~\ref{sec:motor}).
The objective of the implementation is to validate the feasibility of our detection method in practice.
One of the reasons why we choose these two systems is that they are widely deployed in many critical applications: the speaker system can be found in applications in which sound information needs to be broadcast, such as mobile phones and car satellite navigation; the motor control system can be found in those that need to drive some mechanical structures, such as smart locks and insulin pumps.
Implementing the detection method on these two systems also verifies its capabilities of handling different actuator systems regardless of types of signals:
sinusoidal signals (analog) are used in the speaker system, while pulses (digital) in the motor control system.

We first introduce how to build our own actuator systems on which we can quickly implement our detection method. 
Then, we show how to detect various attacking signals in each actuator system.
We only demonstrate the control signal injection, as the drive signal injection is power-consuming and difficult to achieve with our equipment (please see detailed discussion in Section~\ref{sec:difficulty_of_drive signal_injection}).
Finally, a summary of the implementation of these two actuator systems is given in Section~\ref{sec:implementation_summary}.

\subsection{Setup}
\label{sec:setup}

\begin{figure}[t]
	\def\svgscale{0.55}
	\centering
    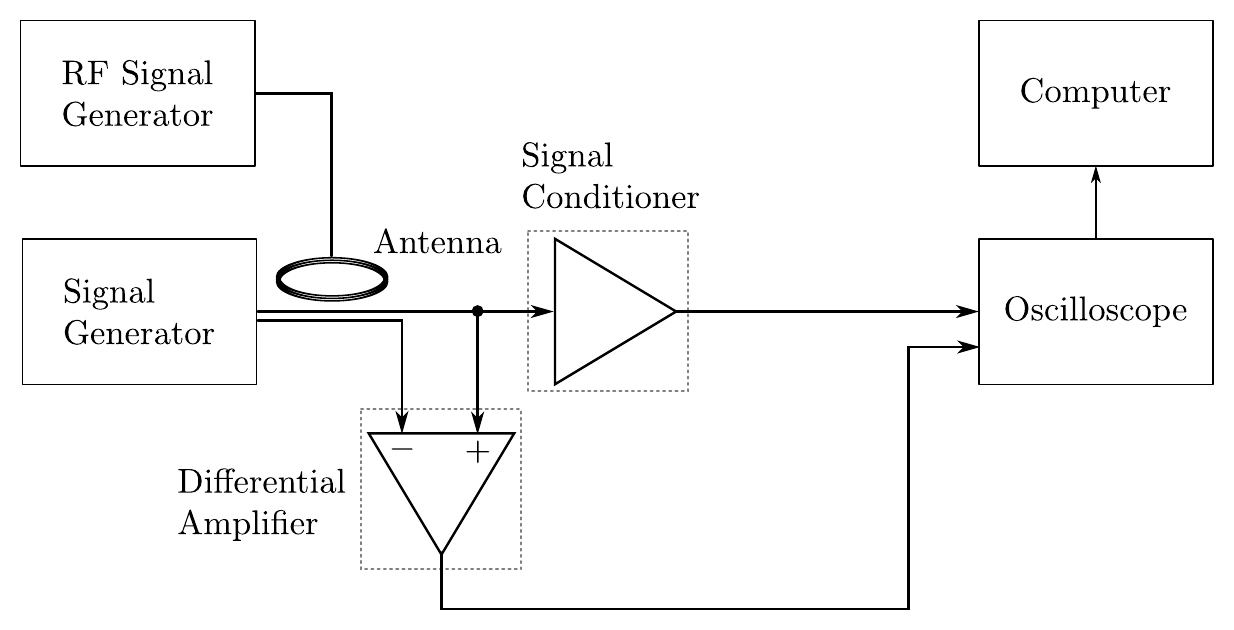
    \caption{A setup of the actuator system. Devices in the dotted squares differ from system to system, and others are the same.}
    \label{fig:speaker_system}
\end{figure}

Based on the system model, we build a setup that can be easily configured into a speaker system or a motor control system, as shown in Figure~\ref{fig:speaker_system}.
We use a signal generator to produce the control signal and the reference signal.
The signal generator is functionally equivalent to the microcontroller.
The benefit of using the signal generator is having easier control of signals regarding their frequencies, amplitudes, synchronization, etc.

The control signal is fed into a signal conditioner.
The signal conditioner is different in these two systems: an audio power amplifier LM386 is used in the speaker system, and a brushed DC motor driver chip DRV8833 is used in the motor control system.

Regarding the actuator (either a loudspeaker or a motor), since its responses are deterministic and its input signal (i.e., the drive signal) sufficiently reflects the responses, we simply omit the actuator in the setup but use an oscilloscope to monitor and record the drive signal.
An advantage of doing so is that different actuator systems can be quickly tested without extra work of using different methods to sense and process the actuator responses (e.g., a microphone to measure sound played by the speaker, or a hall-effect sensor to measure the speed of the motor).
Moreover, a computer is used to process the data that is recorded by the oscilloscope.

Based on such an actuator system, we deploy our detection method to it.
The control signal and the reference signal are fed into a differential amplifier, as shown in Figure~\ref{fig:speaker_system}.
In the speaker system, we choose an AD623 with a gain of around 150 as the differential amplifier because it is specifically designed to amplify small differences between its two inputs.
As for the motor control system, a unity-gain differential amplifier AD629 is selected as the differential amplifier, as it can handle high-voltage inputs.
The output of the differential amplifier is monitored and recorded by the oscilloscope, and the recorded data are sent to the computer for attack detection.

To achieve a large $K$, i.e., difference between the transfer functions of the control signal wire and the reference wire, we form a loop on the control signal wire to make it easier to pick up the attacking signal, and choose a short cable as the reference wire.
Thus, the control signal wire is much more sensitive to the attacking signal than the reference wire.
Note that it does not matter which wire is more sensitive because our detection method only requires the transfer functions to be different. 
Moreover, to guarantee that the control signal and the reference signal arrive at the differential amplifier at the same time, the tapping point is carefully chosen to ensure that the paths that feed these two signals into the differential amplifier have the same length.

Our setup is extremely flexible and allows us to easily experiment with different actuator types without having to build dedicated systems for each one. Despite being a lab setup we believe that our results accurately reflect the response of real commercial products.

\subsection{Speaker System}
\label{sec:speaker}
In a speaker system, an audio signal is amplified and then broadcast. 
The objective of the attack is maliciously manipulating the waveform of the audio signal, and in the extreme case, can lead to the speaker system broadcasting any messages the attacker wishes.

\subsubsection{Determining Threshold}

The differential amplifier output is measured when no attack happens. 
The measurements show that the differential amplifier output signal amplitude is always below $\SI{2.4}{\milli\volt}$.
Since this value already includes all noise sources in our experimental environment, it is chosen as the threshold. 
The benefit of choosing this value as the threshold is that, on the one hand, it significantly reduces the possibility of which the noise accidentally triggers the detection; the false-positive rate remains at $0\%$ as calculated from the measurements.
On the other hand, this threshold is small enough to guarantee that the weakest attack that effectively impacts the actuator system is successfully detected, and this will be shown and explained in the experimental results as follows.

\subsubsection{Direct Power Injection Attacks}
\label{sec:speaker_in-band_detection}
\begin{figure}[t]
	\def\svgscale{0.55}
	\centering
    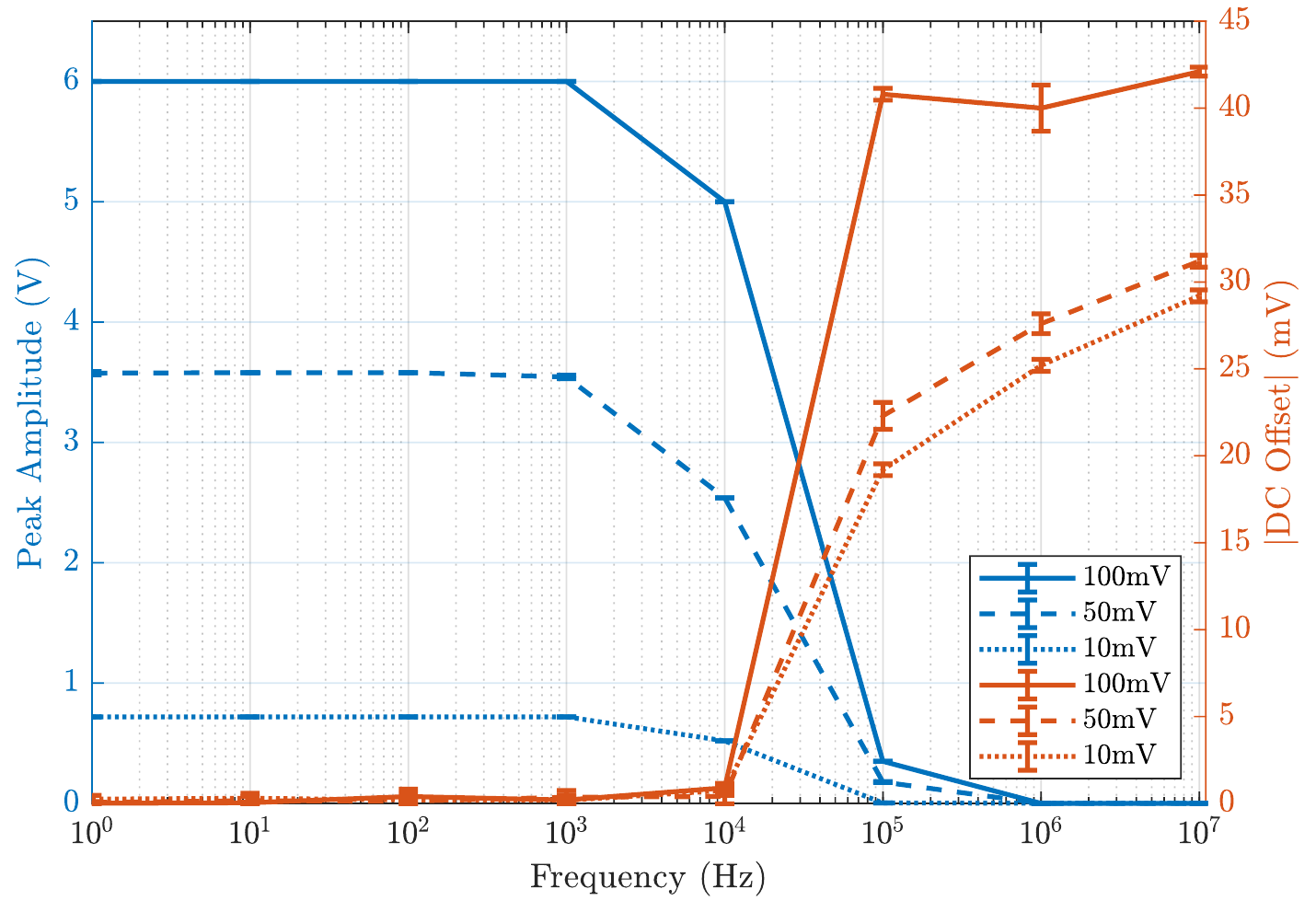
    \caption{The peak amplitude (left y-axis) of the differential amplifier output drops to zero when the frequency of the attacking signal is far beyond the operational band of the audio amplifier; the DC offset (right y-axis) rises while increasing the frequency of the attacking signal. The peak-to-peak voltage of the attacking signal is from $\SI{10}{\milli\volt}$ to $\SI{100}{\milli\volt}$}
    \label{fig:speaker_comparator_output_dpi}
\end{figure}
The normal operational band of an audio amplifier is below the megahertz level, and low-frequency attacking signals are needed for in-band attacks.
Due to the practical difficulty of injecting low-frequency attacking signals into the circuit wirelessly, we first demonstrate that the detection method can handle the in-band attacks using direct power injection (DPI)~\cite{giechaskiel2019sok}.
Note that in the following sections (Section~\ref{sec:speaker_out-of-band_detection} and Section~\ref{sec:motor_detection}), the attacking signals are injected wirelessly.

In order to show that any malicious frequency can be injected into the audio signal, the attack frequency is swept from $\SI{1}{\hertz}$ to $\SI{10}{\mega\hertz}$, and the peak-to-peak voltage of the attacking signal is from $\SI{10}{\milli\volt}$ to $\SI{100}{\milli\volt}$.
The reason why the highest attack frequency is set to $\SI{10}{\mega\hertz}$, which is beyond the operational band of the audio amplifier, is to verify that no gap (as described in Section~\ref{sec:protection_outofthe_operational_band}) exists in the frequency band.
The reason why $\SI{10}{\milli\volt}$ is chosen as the weakest peak-to-peak amplitude of the attacking signal, is that the malicious change caused by an attack at this voltage is already around $\SI{49}{\dB}$ weaker than the audio signal. Weaker attacking signals have little to no impact on the speaker system.

To demonstrate the impact of the attack on the differential amplifier in detail, we show both the peak amplitude and the DC offset in Figure~\ref{fig:speaker_comparator_output_dpi}.
Each point in the figure represents the averaged peak amplitude or the averaged DC offset with a standard deviation.
The first observation of the experimental results is related to the attack power: the peak amplitude and the DC offset increase (decrease) while the attack power increases (decreases).
Concerning the attack frequency, when it is lower than $\SI{1}{\kilo\hertz}$, the peak amplitude is significantly larger than the threshold, which reveals the existence of the attacking signal.
When the frequency is between $\SI{1}{\kilo\hertz}$ and $\SI{1}{\mega\hertz}$, the peak amplitude plummets, but it is still above the threshold; meanwhile, the DC offset rises above the threshold.
When the frequency of the attacking signal reaches $\SI{1}{\mega\hertz}$ and beyond, the DC offset is well above the threshold, indicating the existence of the attack.

The experimental results validate the capabilities of the differential amplifier to detect attacks in the entire frequency range from DC to $\SI{10}{\mega\hertz}$.
We perform this experiment 240 times and all (240 out of 240) attacking signals are detected, making the true-positive rate is $100\%$.
This shows that even for practical systems, the detection method provides strong protection against both in-band and out-of-band attacks.

\subsubsection{Wireless Attacks}
\label{sec:speaker_out-of-band_detection}

\begin{figure}[t]
	\def\svgscale{0.55}
	\centering
    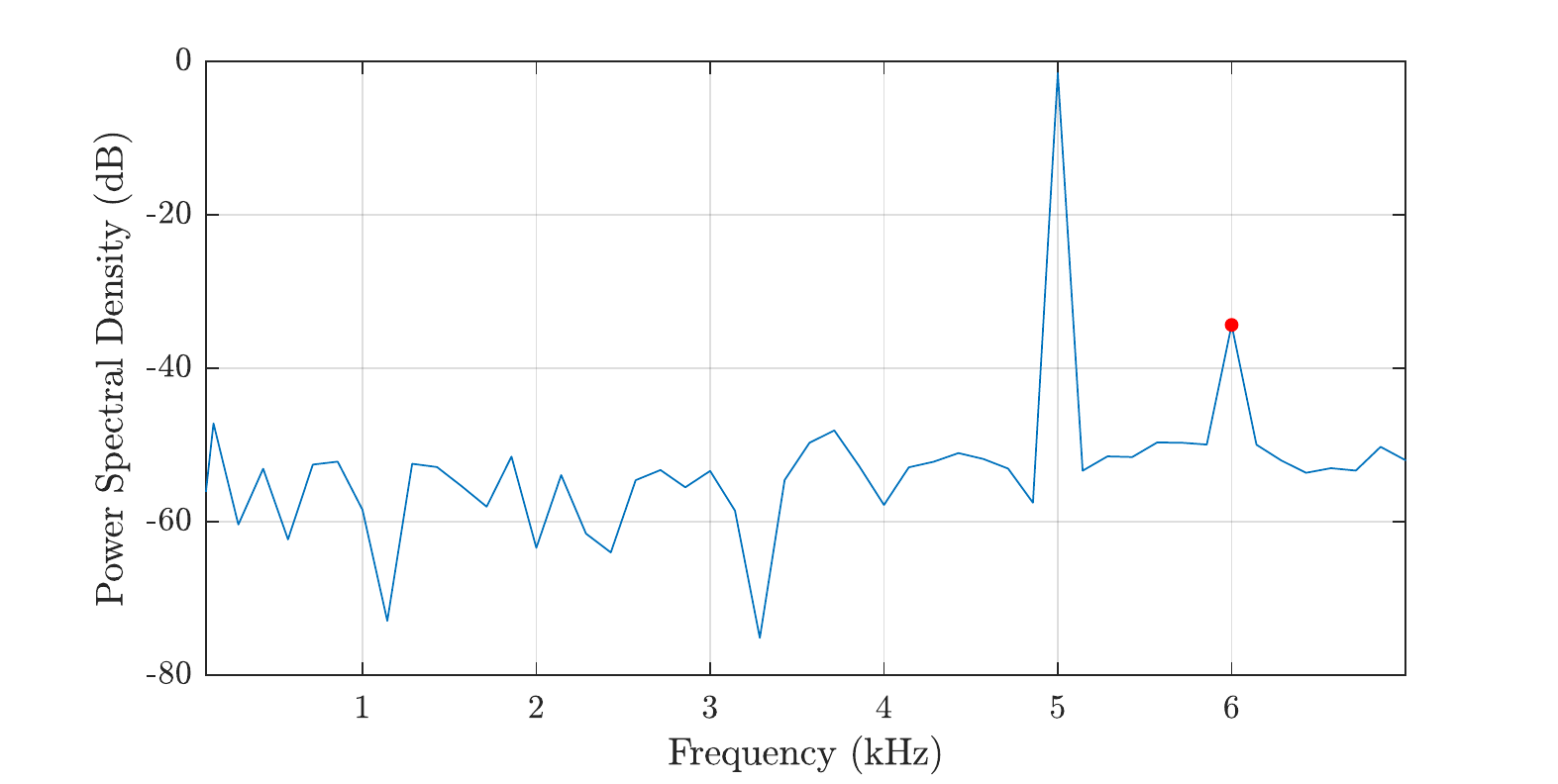
    \caption{A $\SI{6}{\kilo\hertz}$ malicious signal is successfully injected into the $\SI{5}{\kilo\hertz}$ audio signal. The $\SI{6}{\kilo\hertz}$ spike is highlighted by a red point in the frequency domain of the audio amplifier output. The power ratio between the $\SI{6}{\kilo\hertz}$ frequency component and the $\SI{5}{\kilo\hertz}$ is around $\SI{-30}{\dB}$.}
    \label{fig:speaker_amp_output_DPI}
\end{figure}

To test high-frequency attacks in a more realistic setting, we modulate a high frequency carrier with an audio signal and inject it wirelessly into the control and reference wires.
An RF signal generator is used to produce the attacking signals, and they are radiated by a coil antenna, as shown in Figure~\ref{fig:speaker_system}. 
The antenna is placed around $\SI{2}{\centi\meter}$ above the control signal wire for the best possible energy transfer. 
That way we can use less power to achieve the wireless attack in our experiments.
If an attacker is further away from the victim system, she needs more powerful attacking signals to achieve the attack. 

To present a concrete attack, we choose to inject a $\SI{6}{\kilo\hertz}$ malicious frequency into a $\SI{5}{\kilo\hertz}$ audio signal.
In Figure~\ref{fig:speaker_amp_output_DPI}, an attack result is shown: in the frequency domain of the audio amplifier output, besides the legitimate $\SI{5}{\kilo\hertz}$ frequency component, a malicious spike can be observed at $\SI{6}{\kilo\hertz}$.
In order to quantify the impact of the attack, the power ratio between the malicious frequency component and the legitimate frequency component is measured, which can be expressed as the following equation:
\begin{equation*}
impact = 10\times\log_{10}\Big(\frac{P_{malicious}}{P_{legitimate}}\Big)
\end{equation*}
where $P$ represents the power.
The bigger the ratio is, the stronger the injected signal is, and the larger the impact of the attack is. 
When no attacking signal is presented, our measurements show that the $impact$ remains at around~$\SI{-52.7}{\dB}$.

\begin{figure}[t]
	\def\svgscale{0.55}
	\centering
    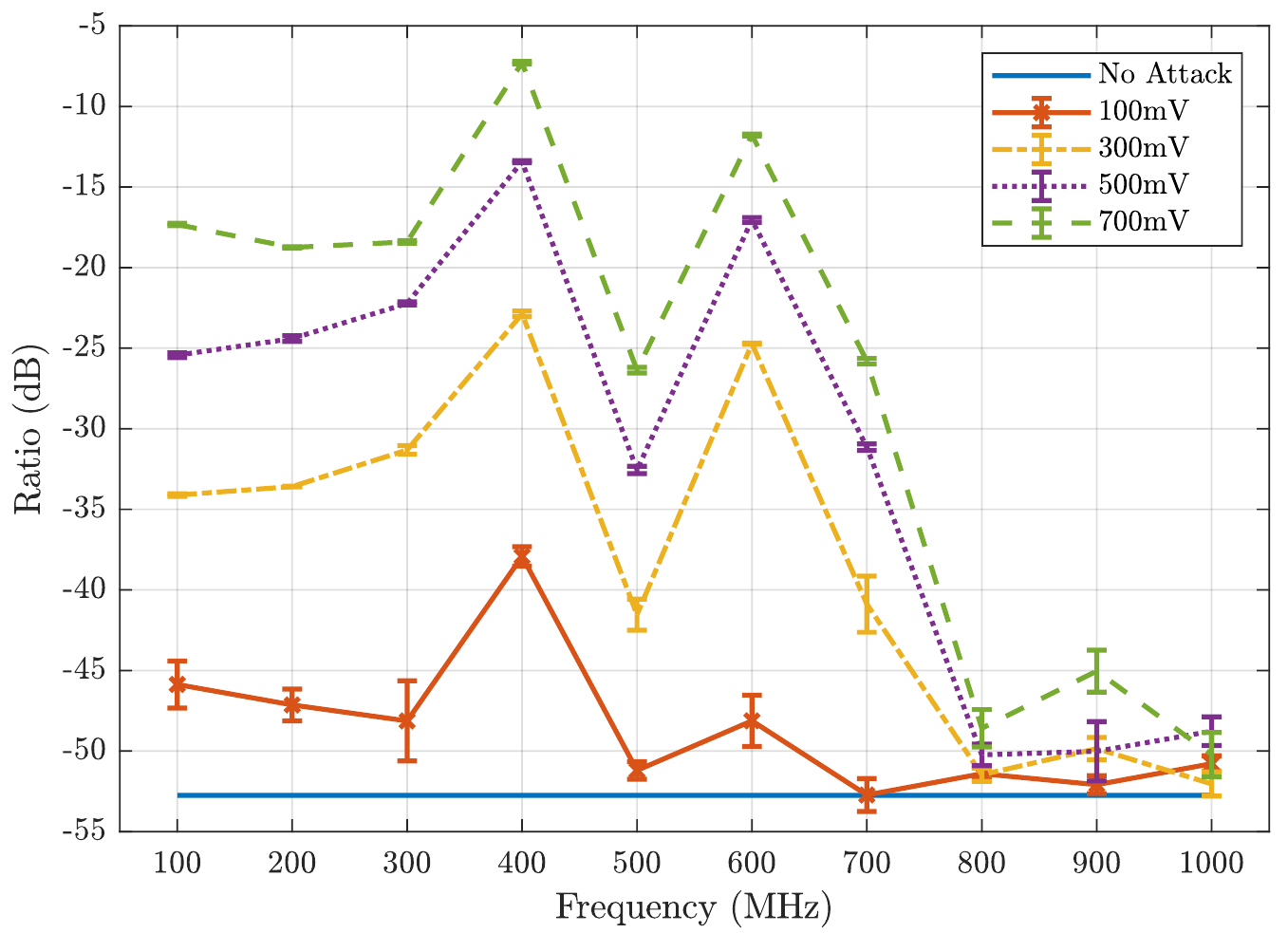
    \caption{The power ratio between the malicious signal and the legitimate signal gradually decreases while increasing the frequency of the attacking signal. The peak-to-peak voltage of the attacking signal is from $\SI{100}{\milli\volt}$ to $\SI{700}{\milli\volt}$.}
    \label{fig:speaker_amp_output_attack_ratio}
\end{figure}

Different attacking signals are generated to test the performance of the detection method:
the peak-to-peak voltage of the attacking signal is changed from $\SI{100}{\milli\volt}$ to $\SI{700}{\milli\volt}$, and the carrier frequency of the RF signal is changed from $\SI{100}{\mega\hertz}$ to $\SI{1000}{\mega\hertz}$.
The impact of the attacks are numerically represented in Figure~\ref{fig:speaker_amp_output_attack_ratio}.
When the attack frequency reaches $\SI{800}{\mega\hertz}$, the $impact$ is close to $\SI{-52.7}{\dB}$, which means that attacks beyond this frequency will have little practical significance.
We did conduct experiments beyond $\SI{1000}{\mega\hertz}$, but the impact of the attacking signal beyond $\SI{1000}{\mega\hertz}$ are smaller, and hence we only focus on the frequency range within $\SI{1000}{\mega\hertz}$.

\begin{figure}[t]
	\def\svgscale{0.55}
	\centering
    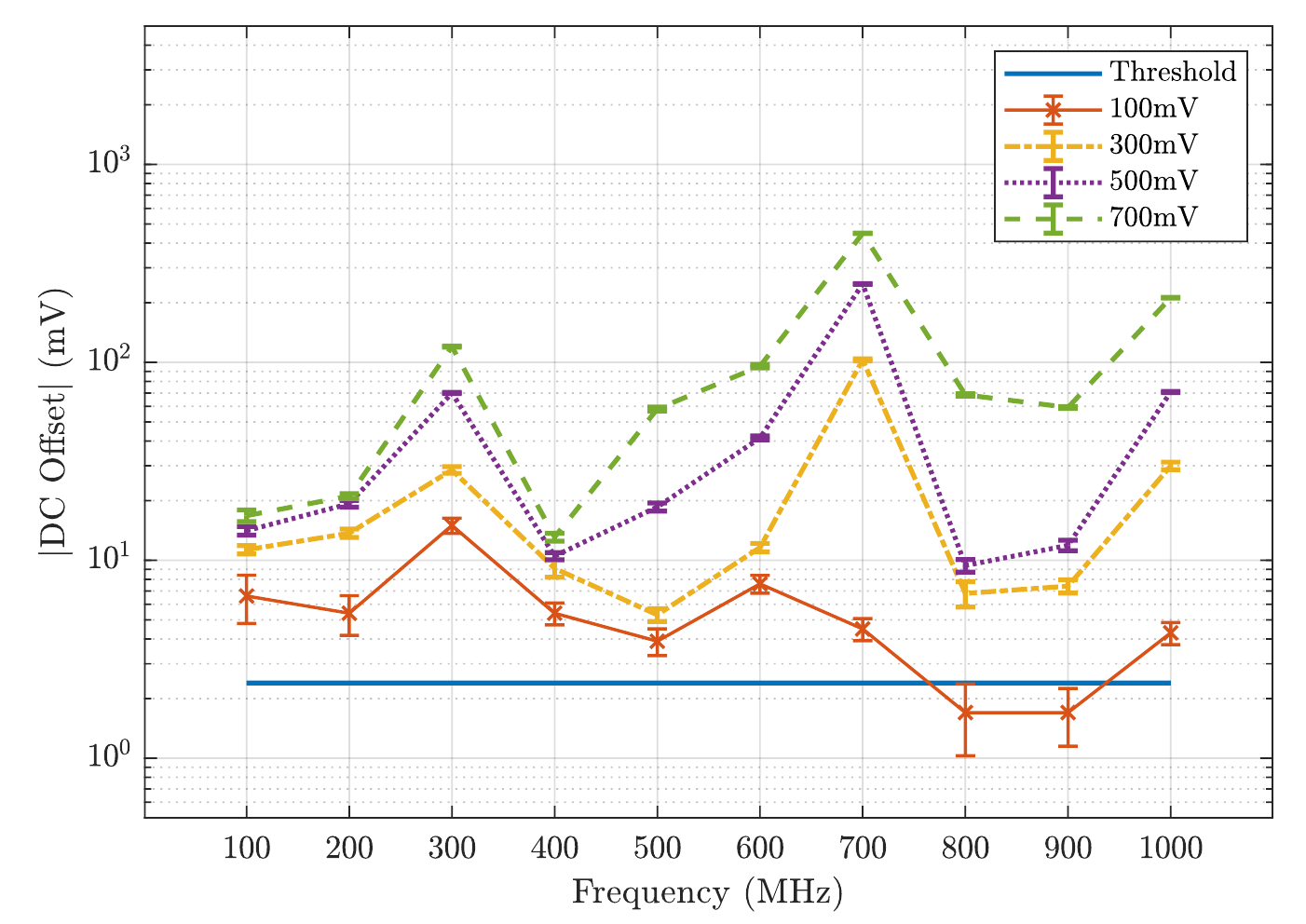
    \caption{The DC offset of the differential amplifier output varies while changing the voltage level and the frequency of the attacking signal. The peak-to-peak voltage of the attacking signal is from $\SI{100}{\milli\volt}$ to $\SI{700}{\milli\volt}$.}
    \label{fig:speaker_comparator_output}
\end{figure}

Regarding the attack detection, the peak amplitudes of all measurements of the differential amplifier output are below the threshold. 
This is because the frequency of the attacking signal is already far beyond the operational band of the differential amplifier, as explained in Section~\ref{sec:protection_outofthe_operational_band}.
However, as shown in Figure~\ref{fig:speaker_comparator_output}, the DC offset of the differential amplifier output is well above the threshold throughout the range for all attacker signals other than  $\SI{100}{\milli\volt}$, indicating the existence of the attack.
We see that the DC offset increases when the attack power is increased, so for attacking signals with peak-to-peak voltages of $\SI{300}{\milli\volt}$, $\SI{500}{\milli\volt}$, and $\SI{700}{\milli\volt}$, the DC offsets are always above the threshold (solid blue line) regardless of the frequency.
When the attacking signal is $\SI{100}{\milli\volt}$, a few attacks fall below the threshold when the carrier frequencies reach $\SI{800}{\mega\hertz}$ and $\SI{900}{\mega\hertz}$.
Referring back to the impact of these two attacking signals in Figure~\ref{fig:speaker_amp_output_attack_ratio}, the ratios indicate that the impacts are so tiny that they are unlikely to have any significance for a practical system.
Since our detection method successfully detects 389 out of 400 attacking signals, the true-positive rate is $97.25\%$.

In Figure~\ref{fig:speaker_comparator_output}, the curves of DC offsets vary up and down along the attack frequency.
This is because the attacking signal is injected wirelessly instead of through DPI.
The transfer function of the wire accounts for the ups and downs of the curves: the attacking signal is efficiently injected into the wire at specific frequencies where local maximum values of the DC offset reaches, but less efficient at other frequencies.

The experiment results show that the frequency range covered by the differential amplifier is easily large enough to protect the frequency band that the speaker system is vulnerable to.
Our detection method shows the feasibility of detecting the attacking signals with frequencies from DC to far beyond the speaker system's operational band.
Moreover, given the wireless injections, our detection method demonstrates its capabilities of handling real attack scenarios.
We present concrete attacking signals that can precisely manipulate the audio frequencies, but it does not mean that our detection method can only handle these specific attacking signals.
Any attacks that cause voltage changes of the differential amplifier output signal beyond the pre-determined threshold can be spotted immediately.

\subsection{Motor Control System}
\label{sec:motor}

\begin{figure}[t]
	\def\svgscale{0.60}
	\centering
    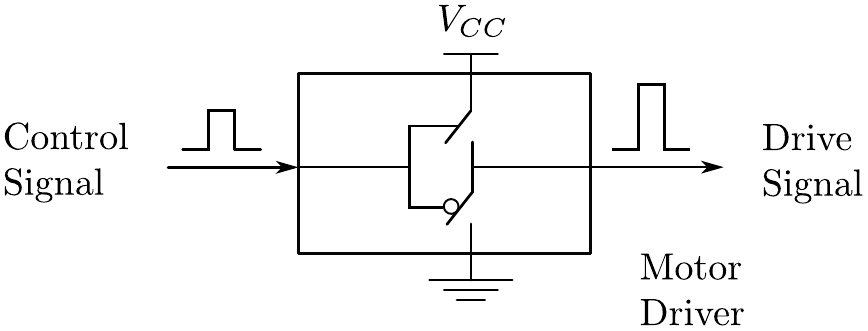
    \caption{A motor driver is used to amplify a control signal to drive a motor.}
    \label{fig:experiment_setup}
\end{figure}

In the motor control system, a pulse signal is used to control the rotating speed of the motor.
The duty cycle of the pulse signal describes the amount of time that the signal is at the high-voltage level as a percentage of the total time of a cycle.
The larger the duty cycle is, the faster the motor's rotation speed is.
As mentioned in the setup, a motor driver is used as a signal conditioner to amplify the control signal into a powerful drive signal to energize the motor.
The motor driver is made of transistors, and for simplicity, as shown in Figure~\ref{fig:experiment_setup}, they can be regarded as two switches that are connected in series and are controlled by the pulse signal.
Since these two switches work in opposite ways, the output signal toggles between $V_{CC}$ and the ground in the same pattern as the input signal.
The attacker's objective is to manipulate the duty cycle and impact the functionality of the motor.


\subsubsection{Determining Threshold}
\label{sec:motor_setup}
When no attack presents, the differential amplifier output signal is recorded, and the threshold is $\SI{0.17}{\milli\volt}$.
This threshold value is chosen as it makes the false-positive rate to its minimum ($0\%$) in our experimental environment; also, this threshold is sufficiently large to spot the weakest attacks, as shown as follows.

\subsubsection{Detection of Attacks}
\label{sec:motor_detection}

\begin{figure}[t]
	\def\svgscale{0.55}
	\centering
    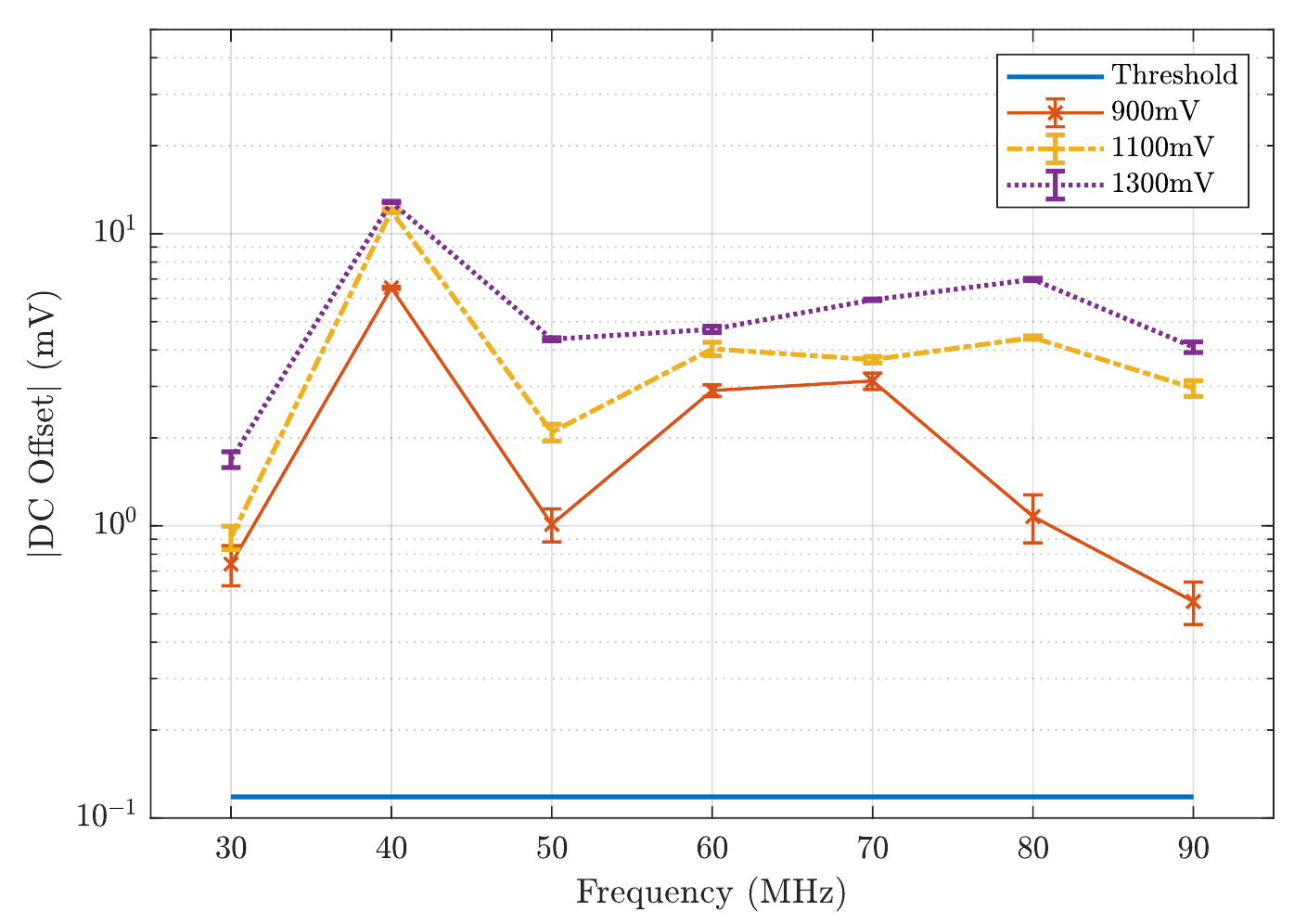
    \caption{When an attack happens, the DC offset of the differential amplifier output is always above the threshold, implying detecting the attack.}
    \label{fig:motor_comparator_output}
\end{figure}

Since the differential amplifier is specifically designed to handle the input difference in its operational band, it is not difficult to detect the in-band attacks.
We do not repeat the in-band attacks here but focus on the out-of-band attacks.
Note that the out-of-band attacks are realized wirelessly.

In the experiments, the frequency of the attacking signal ranges from $\SI{30}{\mega\hertz}$ to $\SI{90}{\mega\hertz}$, and the peak-to-peak voltage ranges from $\SI{900}{\milli\volt}$ to $\SI{1300}{\milli\volt}$.
The reason why the frequency of the attacking signal is below $\SI{90}{\mega\hertz}$ is that beyond this frequency, the motor driver never responds to the attacking signal, even though the peak-to-peak voltage of the attacking signal reaches its upper limit in the signal generator. 
The reason why the peak-to-peak voltage of the attacking signal is above $\SI{900}{\milli\volt}$ is that, below this voltage level, the attacking signal is too weak to affect the motor driver.
In our experiment, the RF signals can cause the motor driver to output a low voltage level when a high voltage level should be outputted.
In other words, the duty cycle of the pulses can be reduced.
We can precisely control when to start and stop radiating the attacking signals, and hence, the duty cycle of the control signal can be precisely manipulated, further controlling the motor speed.
Note that using other types of attacking signals can also increase the duty cycle~\cite{selvaraj2018electromagnetic}; however, the purpose of the experiment focuses on attack detection, and we do not further show and discuss how to control the motor speed.

Regarding the attack detection, both the peak amplitude and the DC offset of the differential amplifier output signal are checked. 
Under these out-of-band attacks, the peak amplitude is always below the threshold. 
However, as shown in Figure~\ref{fig:motor_comparator_output}, the DC offset is always above the threshold, indicating an attack.
All (210 out of 210) DC offsets are above the threshold, indicating that all attacking signals are detected.
Therefore, the true-positive rate is $100\%$.

\subsection{Summary of Implementation}
\label{sec:implementation_summary}
The implementation of our detection method on the speaker system and the motor control system show the generality of our detection method regardless of the type of signal.
The deployments also demonstrate the simplicity of implementing the detection method in practice.
The high true-positive rates and low false-positive rates in the speaker system and the motor control system show the robustness of the detection method on different actuator systems.

\section{Discussion}
\label{sec:discussion}


\subsection{Different Detection Strategies}
\label{sec:strategies}

\begin{figure}[t]
  \def\svgscale{0.60}
  \centering
  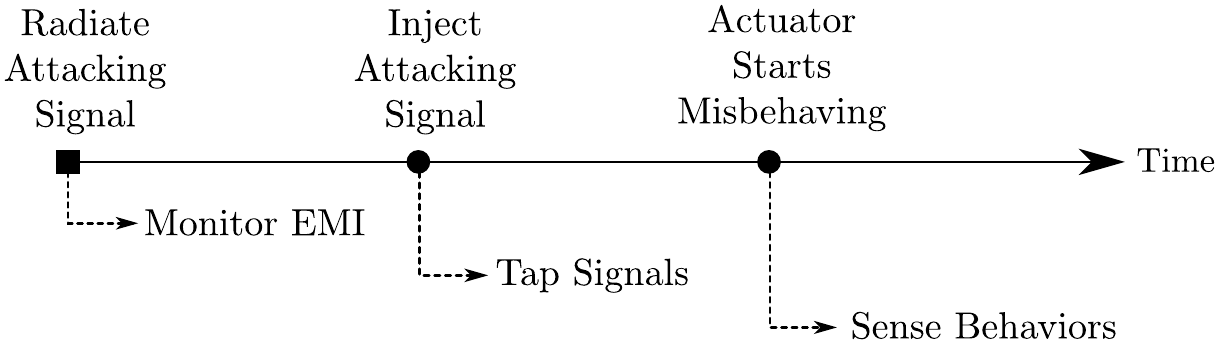
  \caption{The timeline shows that an attack starts from radiating an attacking signal to actuator misbehaving. At different moments, the attack can be detected by different ways.}
  \label{fig:timeline}
\end{figure}

An electromagnetic signal injection attack has several distinct phases, each of which gives rise to different detection strategies. 
In the attack timeline shown in Figure~\ref{fig:timeline}, three moments are highlighted:
the first moment is when the attacking signal is radiated; the second is when the attacking signal is injected into a wire in the target system; the third is when the actuator starts deviating from its intended activity.
Each of the three moments marks the start of a new phase of the attack. 
The three phases should not be thought of as sequential since an attack of any meaningful duration will be in all three phases at once, but should rather be thought of as three opportunities to detect the attack.

In the first phase, when attacking signals are being radiated, the electromagnetic radiation can be detected in the environment using an antenna.
Thus, the detection strategy is monitoring the environmental electromagnetic power level: if the power level is above a pre-defined threshold, or maybe outside a known noise profile, the attack is detected. This strategy has the potential to detect an attack early; however, it requires a monitoring device that can reliably detect adversarial interference at the frequency that would harm the target system, which is not always easy to achieve.
Examples of this detection strategy are some of the anomaly detectors that will be introduced in Section~\ref{sec:related_work_detection_anomaly}.

In the second phase, when the attacking signals are successfully injected into the actuator system, the signals in the actuator system wires are changed.
Since the attacking signals are not supposed to exist in the actuator system, these changes are a reliable indicator of an attack, if they can be measured.
Our detection method uses the second form of detection, i.e., we detect signals that are successfully injected into the actuator system. 

In the final phase, the actions of the actuator will deviate from what the system expects, assuming the attack is powerful enough to result in a measurable change. If the system can detect this behavior change, this can be used to detect the attack. This might be an attractive detection strategy since no effort is wasted on attacks that do not have a measurable effect on the target actuator.
However, detecting such attacks typically requires extra sensors.
By the nature of the detection method, it will only detect attacks after they have already affected the system. 
An example of this detection method is Muniraj and Farhood's work~\cite{muniraj2019detection} that will be introduced in Section~
\ref{sec:related_work_detection_actuator}.


\subsection{Adaptive Threshold}
\label{sec:adaptive_threshold}

In our implementation of the detection method, we find a proper threshold by experiment and keep it constant while testing the performance of our detection method.
The advantage of using a constant threshold is that once a proper threshold is found and determined, it is efficacious forever, and the designer never has to adjust it again.
However, in some cases where the environmental noise varies significantly and complicatedly over time, to provide the actuator system with more flexibility, the designer can program the actuator system to adjust its threshold adaptively when necessary.
Imagine a simple case: during the daytime, the noise is intense because of human activities (e.g., wireless communications, transportations), but at midnight when people sleep, the noise becomes relatively weak.
During the daytime, the threshold can be slightly increased to allow more noise, and as such, it can avoid the noise frequently triggering the detection.
At midnight, to restore the detection method to be more sensitive to attacks, the designer can program the actuator system to lower down the threshold.
No matter how the designer adjusts the threshold adaptively, it is still essential to guarantee that the detection method meets the requirements as mentioned in previous sections: first, no noise triggers the detection accidentally; second, no attack that effectively impacts the actuator is missed.

\subsection{Difficulty of Canceling Attacking Signals}

An idea of mitigating the influence caused by attacks is generating an ``anti-attack'' signal to cancel out the attacking signal.
The anti-attack signal and the attacking signal have the same frequency and amplitude, but they are 180 degrees out of phase.
When the anti-attack signal and the attacking signal meet, they destruct each other.
This idea is similar to the sound noise cancellation technology that is used in headphones. 
However, it is hard to realize such a cancellation regarding the electromagnetic interference.
In the air, an electromagnetic signal propagates around the light speed; in the circuit, the speed halves.
In addition, it takes time for the actuator system to capture the attacking signal and then generate the anti-attack signal for the cancellation.
This means that the anti-attack signal always lags behind the attacking signal.
It is difficult to synchronize the anti-attack signal with the attacking signal unless the microcontroller can predict the attacking signal.

\subsection{Difficulty of Drive Signal Injection}
\label{sec:difficulty_of_drive signal_injection}

As mentioned previously, compared with the control signal injections, a drive signal injection may require much more power if the actuator is power-consuming.
We estimate the power of a drive signal injection as follows.
According to datasheets of an off-the-shelf motor, it needs a drive signal that is around $\SI{4.5}{\watt}$; as for a microcontroller, such as an Arduino Uno microcontroller, it can output a control signal that is only $\SI{0.1}{\watt}$.
For simplicity, we suppose that the attenuation on attacking signals is the same in those two injections.
Then, the attacker needs to radiate at least $\frac{\SI{4.5}{\watt}}{\SI{0.1}{\watt}}=45$ times more power to realize the drive signal injection than the control signal injection.
This result implies that it is much more difficult and costly to conduct the drive signal injection than the control signal injection in practice.

Another evidence to show that the drive signal injection is hard to achieve is to regard the injection as wireless power transmission~\cite{shinohara2014wireless}. 
In wireless power transmission techniques, scientists specifically designed both antennas of the transmitter and the receiver to achieve the power transmission.
Given the wire that works as a low-gain antenna in the actuator system, delivering enough power into the drive signal wire can be much more challenging.

\section{Related Work}
\label{sec:related_work_existing_defenses}

Many countermeasures against electromagnetic signal injection attacks have been proposed and developed; however, it needs to be noted that protecting sensors has been much more extensively studied than actuators.
The countermeasures can be categorized into two types: one is attenuation that aims to reduce attack impacts, and the other is detecting the existence of attacks. 

\subsection{Attenuation}
\label{sec:related_work_attenuation}

Wrapping components with proper RF shielding materials is a common method to attenuate attacking signals~\cite{kasmi2015iemi,kune2013ghost,
Kasper2009PACf,selvaraj2018electromagnetic,
Markettos2009Tfia, osuka2018information,giechaskiel2019framework, shin2016sampling, tu2019trick, sp20Zhang,wang2022ghosttouch}.
However, the shielding materials provide finite attenuation~\cite{tesche1996emc}, and a powerful attacker may breach the protection by increasing her attack power.
Although adding thicker shielding materials can increase the attenuation level, it will still challenge the weight and the size of the devices, especially for applications such as implantable medical devices and aviation.
In addition to shielding materials, regarding traces in a printed circuit board (PCB), researchers suggested that via-fenced striplines can also eliminate attacking signals by a finite amount~\cite{dayanikli2020senact, dayanikli2021electromagnetic}.

Filtering is another prevalent solution to mitigate attacking signals.
Low-pass filters can significantly attenuate out-of-band attacking signals~\cite{giechaskiel2019framework, selvaraj2018electromagnetic, kune2013ghost, tu2019trick,osuka2018information, sp20Zhang}.
However, in-band attacking signals can still pass through the low-pass filters.
Researchers also pointed out that the parasitics in
surface mount components can convert the low-pass filter into a band-stop filter, which allows out-of-band attacking signals to pass~\cite{ryanhurley2007}.
Besides, Kune et al.~\cite{kune2013ghost} proposed to deploy an adaptive filtering mechanism~\cite{proakis2001digital} that makes use of knowledge about ambient electromagnetic emissions to attenuate the interference in sensor measurements.
Crovetti and Musolino~\cite{crovetti2021digital} also proposed a digital way to suppress the EMI-induced errors in the sensor measurements.
However, it is challenging to have such digital methods for the actuator because it has no computational capabilities.
Furthermore, Kune et al.~\cite{kune2013ghost} also recommended using differential rather than single-end comparator to attenuate the attacking signals in a finite frequency band, thereby raising the bar for attackers.

\subsection{Detection}
\label{sec:related_work_detection}

\subsubsection{Anomaly Detection}
\label{sec:related_work_detection_anomaly}
An idea of detecting attacks is to add a specific channel to monitor whether abnormal electromagnetic signals or activities appear.
Note that although some of the following approaches are initially designed for sensors, similar ideas possibly work in actuator systems, too. 
Researchers developed standalone detection systems that capture electromagnetic waves by dedicated antennas and then use intricate circuits to process the captured signals for detection~\cite{adami2014hpm,adami2011hpm,dawson2014cost}.
Kune et al.~\cite{kune2013ghost} investigated using extra antennas or conductors to capture and measure attacking signals for detection, and the measurements can be then used by their adaptive filtering mechanism as mentioned previously.
In a similar vein, Tu et al.~\cite{tu2021transduction} proposed adding a dummy sensor for detection and correction.
In another work, Tu et al.~\cite{tu2019trick} proposed leveraging the superheterodyne technique to create an anomaly detector to check whether sensor measurements carry malicious frequency components.
Note that these approaches count on the knowledge about the waveforms of the attacking signals, which are usually high-frequency (e.g., MHz or GHz). 
Thus, they require electronic components (e.g., high-speed ADCs) that can properly handle high-frequency signals, as well as extra computing resources to process the captured signals for detection purposes, implying significant implementation overheads regarding both hardware and software.

As a comparison, our approach counts on the signal strength difference between the primary and the reference signals to detect attacks, rather than waveforms. 
This makes our approach gain advantages over the other approaches: first, a simple detection circuit made of differential amplifiers is used to catch the difference, and such a detection circuit has fewer hardware overheads; second, an interrupt pin of the microcontroller is configured to handle the output of the detection circuit to determine whether an attack happens, and it needs fewer computing resources. 
Besides, our approach does not require any RF interface to capture the attacking signals, thus avoiding the troubles of crafting the dedicated RF interfaces, as well as preventing extra attack power from entering the victim devices and causing other unwanted influence.

\subsubsection{Detection Methods for Sensor Systems}
\label{sec:related_work_detection_sensor}
Especially for sensor systems, Zhang and Rasmussen~\cite{sp20Zhang} proposed a generalized detection method that selectively turns off the sensor in a secret way to observe whether attacks alter the sensor measurements. 
Shoukry et al.~\cite{shoukry2015pycra} proposed similar detection methods, but they were designed for specific types of sensors, as well as requiring significant computational overheads.
Succeeding studies~\cite{kohler2021signal, ruotsalainen2021watermarking} further adapted these detection methods to more practical applications.
Fang et al.~\cite{fang2022detection} proposed adding unique noise (fingerprints) to sensor measurements and using machine learning techniques to detect the attacks.

In addition, in specific devices such as cardiac implantable electrical devices (CIED)~\cite{kune2013ghost} and smartphones~\cite{wang2022ghosttouch}, researchers utilized users' reactions or behaviors while using these devices to identify the existence of attacks on the sensors.
A few works mentioned that multiple built-in sensors of a device can react to variations of the electromagnetic environment, and the characteristics can be exploited to detect abnormal electromagnetic activities~\cite{kasmi2015iemi, kasmi2015automated}.
Such a detection approach is also known as sensor fusion, which has been widely studied to detect signal injections that use other types of attacking signals such as ultrasonics and lasers~\cite{giechaskiel2019sok, yan2020sok}.

These detection methods work well for the sensors because 
the computational capabilities of the receiver (microcontroller) make the authentication possible.
However, it is not easy to apply similar ideas to the actuator systems because the receiver (actuator) lacks computational capabilities to authenticate its input signals.  


\subsubsection{Detection Methods for Actuator Systems}
\label{sec:related_work_detection_actuator}
Reliable sensor measurements can be used to indicate whether actuators are under attack.
In unmanned aircraft systems, Muniraj and Farhood~\cite{muniraj2019detection} proposed to artificially cause minor disturbances to the actuators at a random time and use sensors to capture the disturbances; if unexpected disturbances are detected, the attacks are found.
However, this method trades off the stability of the whole system against its security.
The same authors proposed another detection method that casts the actuator attack detection problem as an unknown input estimation problem and uses a two-stage extended Kalman filter to estimate actuator attacks from sensor measurements, requiring additional computational power.
In addition to the two detection methods, the authors also proposed a method that adds randomness to control signals to improve the resilience of the actuator against malicious attacks.

Our approach outstrips these detection methods in terms of these three aspects.
First, they require a complex model of the specific actuator system, which makes it difficult to be applied to other applications, whereas our approach is generalized for different actuator systems.
Second, they need extra computing resources to run the detection algorithms, but we can use the interrupt mechanism of the microcontroller for detection, which is more efficient.  
Third, their detection methods always spot attacks after the actuator misbehaves; however, our approach detects the attacks earlier, thus possibly allowing the actuator systems to take proper measures to stop/mitigate the attacks.

\section{Conclusion}
\label{sec:conclusion}
In this paper, we have proposed a novel detection method that can detect electromagnetic signal injection attacks on actuator systems. 
This class of systems previously had to rely on physical security measures and signal decay, and had no meaningful security guarantees against a determined adversary. 
Our detection system fills this critical gap and provides strong detection guarantees to any actuator system.
The core idea of our detection method is straightforwad: any diffenrence caused by external attacks between two identical signals (the primary signal and the reference signal) indicates the attacks. 
Our detection method provides provable guarantees against attacks, and can be tuned to any attack power and any amount of environmental noise.
We have shown that our detection method provides the actuator system with a strong security guarantee, and an attacker who attempts to effectively manipulate the actuator system will always be detected by our detection method.
Despite this, our detection method requires only a few cheap off-the-shelf electronic components and does not add any significant weight to the system it protects. 
This is important in many contexts, such as aviation and implantable medical devices. 
Moreover, the implementation of the detection method on a speaker system and a motor control system proves its generality for different actuator systems, as well as the effectiveness and the robustness in a practical setting.

\balance
\bibliographystyle{ACM-Reference-Format}
\bibliography{HackMotors.bib}
\end{document}

%% file: Figures/block/transfer_functions.pdf_tex
\begingroup%
  \makeatletter%
  \providecommand\color[2][]{%
    \errmessage{(Inkscape) Color is used for the text in Inkscape, but the package 'color.sty' is not loaded}%
    \renewcommand\color[2][]{}%
  }%
  \providecommand\transparent[1]{%
    \errmessage{(Inkscape) Transparency is used (non-zero) for the text in Inkscape, but the package 'transparent.sty' is not loaded}%
    \renewcommand\transparent[1]{}%
  }%
  \providecommand\rotatebox[2]{#2}%
  \newcommand*\fsize{\dimexpr\f@size pt\relax}%
  \newcommand*\lineheight[1]{\fontsize{\fsize}{#1\fsize}\selectfont}%
  \ifx\svgwidth\undefined%
    \setlength{\unitlength}{326.24047949bp}%
    \ifx\svgscale\undefined%
      \relax%
    \else%
      \setlength{\unitlength}{\unitlength * \real{\svgscale}}%
    \fi%
  \else%
    \setlength{\unitlength}{\svgwidth}%
  \fi%
  \global\let\svgwidth\undefined%
  \global\let\svgscale\undefined%
  \makeatother%
  \begin{picture}(1,0.49150005)%
    \lineheight{1}%
    \setlength\tabcolsep{0pt}%
    \put(0,0){\includegraphics[width=\unitlength,page=1]{transfer_functions.pdf}}%
  \end{picture}%
\endgroup%

%% file: Figures/block/add_comparator.pdf_tex
\begingroup%
  \makeatletter%
  \providecommand\color[2][]{%
    \errmessage{(Inkscape) Color is used for the text in Inkscape, but the package 'color.sty' is not loaded}%
    \renewcommand\color[2][]{}%
  }%
  \providecommand\transparent[1]{%
    \errmessage{(Inkscape) Transparency is used (non-zero) for the text in Inkscape, but the package 'transparent.sty' is not loaded}%
    \renewcommand\transparent[1]{}%
  }%
  \providecommand\rotatebox[2]{#2}%
  \newcommand*\fsize{\dimexpr\f@size pt\relax}%
  \newcommand*\lineheight[1]{\fontsize{\fsize}{#1\fsize}\selectfont}%
  \ifx\svgwidth\undefined%
    \setlength{\unitlength}{321.04869413bp}%
    \ifx\svgscale\undefined%
      \relax%
    \else%
      \setlength{\unitlength}{\unitlength * \real{\svgscale}}%
    \fi%
  \else%
    \setlength{\unitlength}{\svgwidth}%
  \fi%
  \global\let\svgwidth\undefined%
  \global\let\svgscale\undefined%
  \makeatother%
  \begin{picture}(1,0.82897057)%
    \lineheight{1}%
    \setlength\tabcolsep{0pt}%
    \put(0,0){\includegraphics[width=\unitlength,page=1]{add_comparator.pdf}}%
  \end{picture}%
\endgroup%

%% file: Figures/block/comparator_vs_attacking_signal.pdf_tex
\begingroup%
  \makeatletter%
  \providecommand\color[2][]{%
    \errmessage{(Inkscape) Color is used for the text in Inkscape, but the package 'color.sty' is not loaded}%
    \renewcommand\color[2][]{}%
  }%
  \providecommand\transparent[1]{%
    \errmessage{(Inkscape) Transparency is used (non-zero) for the text in Inkscape, but the package 'transparent.sty' is not loaded}%
    \renewcommand\transparent[1]{}%
  }%
  \providecommand\rotatebox[2]{#2}%
  \newcommand*\fsize{\dimexpr\f@size pt\relax}%
  \newcommand*\lineheight[1]{\fontsize{\fsize}{#1\fsize}\selectfont}%
  \ifx\svgwidth\undefined%
    \setlength{\unitlength}{349.20933443bp}%
    \ifx\svgscale\undefined%
      \relax%
    \else%
      \setlength{\unitlength}{\unitlength * \real{\svgscale}}%
    \fi%
  \else%
    \setlength{\unitlength}{\svgwidth}%
  \fi%
  \global\let\svgwidth\undefined%
  \global\let\svgscale\undefined%
  \makeatother%
  \begin{picture}(1,0.54164219)%
    \lineheight{1}%
    \setlength\tabcolsep{0pt}%
    \put(0,0){\includegraphics[width=\unitlength,page=1]{comparator_vs_attacking_signal.pdf}}%
  \end{picture}%
\endgroup%

%% file: Figures/block/speaker_system.pdf_tex
\begingroup%
  \makeatletter%
  \providecommand\color[2][]{%
    \errmessage{(Inkscape) Color is used for the text in Inkscape, but the package 'color.sty' is not loaded}%
    \renewcommand\color[2][]{}%
  }%
  \providecommand\transparent[1]{%
    \errmessage{(Inkscape) Transparency is used (non-zero) for the text in Inkscape, but the package 'transparent.sty' is not loaded}%
    \renewcommand\transparent[1]{}%
  }%
  \providecommand\rotatebox[2]{#2}%
  \newcommand*\fsize{\dimexpr\f@size pt\relax}%
  \newcommand*\lineheight[1]{\fontsize{\fsize}{#1\fsize}\selectfont}%
  \ifx\svgwidth\undefined%
    \setlength{\unitlength}{355.24783547bp}%
    \ifx\svgscale\undefined%
      \relax%
    \else%
      \setlength{\unitlength}{\unitlength * \real{\svgscale}}%
    \fi%
  \else%
    \setlength{\unitlength}{\svgwidth}%
  \fi%
  \global\let\svgwidth\undefined%
  \global\let\svgscale\undefined%
  \makeatother%
  \begin{picture}(1,0.51067418)%
    \lineheight{1}%
    \setlength\tabcolsep{0pt}%
    \put(0,0){\includegraphics[width=\unitlength,page=1]{speaker_system.pdf}}%
  \end{picture}%
\endgroup%

%% file: Figures/plot/speaker_comparator_output_dpi.pdf_tex
\begingroup%
  \makeatletter%
  \providecommand\color[2][]{%
    \errmessage{(Inkscape) Color is used for the text in Inkscape, but the package 'color.sty' is not loaded}%
    \renewcommand\color[2][]{}%
  }%
  \providecommand\transparent[1]{%
    \errmessage{(Inkscape) Transparency is used (non-zero) for the text in Inkscape, but the package 'transparent.sty' is not loaded}%
    \renewcommand\transparent[1]{}%
  }%
  \providecommand\rotatebox[2]{#2}%
  \newcommand*\fsize{\dimexpr\f@size pt\relax}%
  \newcommand*\lineheight[1]{\fontsize{\fsize}{#1\fsize}\selectfont}%
  \ifx\svgwidth\undefined%
    \setlength{\unitlength}{411.74091725bp}%
    \ifx\svgscale\undefined%
      \relax%
    \else%
      \setlength{\unitlength}{\unitlength * \real{\svgscale}}%
    \fi%
  \else%
    \setlength{\unitlength}{\svgwidth}%
  \fi%
  \global\let\svgwidth\undefined%
  \global\let\svgscale\undefined%
  \makeatother%
  \begin{picture}(1,0.6929718)%
    \lineheight{1}%
    \setlength\tabcolsep{0pt}%
    \put(0,0){\includegraphics[width=\unitlength,page=1]{speaker_comparator_output_dpi.pdf}}%
  \end{picture}%
\endgroup%

%% file: Figures/plot/speaker_amp_output_DPI.pdf_tex
\begingroup%
  \makeatletter%
  \providecommand\color[2][]{%
    \errmessage{(Inkscape) Color is used for the text in Inkscape, but the package 'color.sty' is not loaded}%
    \renewcommand\color[2][]{}%
  }%
  \providecommand\transparent[1]{%
    \errmessage{(Inkscape) Transparency is used (non-zero) for the text in Inkscape, but the package 'transparent.sty' is not loaded}%
    \renewcommand\transparent[1]{}%
  }%
  \providecommand\rotatebox[2]{#2}%
  \newcommand*\fsize{\dimexpr\f@size pt\relax}%
  \newcommand*\lineheight[1]{\fontsize{\fsize}{#1\fsize}\selectfont}%
  \ifx\svgwidth\undefined%
    \setlength{\unitlength}{452.83463583bp}%
    \ifx\svgscale\undefined%
      \relax%
    \else%
      \setlength{\unitlength}{\unitlength * \real{\svgscale}}%
    \fi%
  \else%
    \setlength{\unitlength}{\svgwidth}%
  \fi%
  \global\let\svgwidth\undefined%
  \global\let\svgscale\undefined%
  \makeatother%
  \begin{picture}(1,0.50312989)%
    \lineheight{1}%
    \setlength\tabcolsep{0pt}%
    \put(0,0){\includegraphics[width=\unitlength,page=1]{speaker_amp_output_DPI.pdf}}%
  \end{picture}%
\endgroup%

%% file: Figures/plot/speaker_amp_output_attack_ratio.pdf_tex
\begingroup%
  \makeatletter%
  \providecommand\color[2][]{%
    \errmessage{(Inkscape) Color is used for the text in Inkscape, but the package 'color.sty' is not loaded}%
    \renewcommand\color[2][]{}%
  }%
  \providecommand\transparent[1]{%
    \errmessage{(Inkscape) Transparency is used (non-zero) for the text in Inkscape, but the package 'transparent.sty' is not loaded}%
    \renewcommand\transparent[1]{}%
  }%
  \providecommand\rotatebox[2]{#2}%
  \newcommand*\fsize{\dimexpr\f@size pt\relax}%
  \newcommand*\lineheight[1]{\fontsize{\fsize}{#1\fsize}\selectfont}%
  \ifx\svgwidth\undefined%
    \setlength{\unitlength}{390.69181493bp}%
    \ifx\svgscale\undefined%
      \relax%
    \else%
      \setlength{\unitlength}{\unitlength * \real{\svgscale}}%
    \fi%
  \else%
    \setlength{\unitlength}{\svgwidth}%
  \fi%
  \global\let\svgwidth\undefined%
  \global\let\svgscale\undefined%
  \makeatother%
  \begin{picture}(1,0.72438996)%
    \lineheight{1}%
    \setlength\tabcolsep{0pt}%
    \put(0,0){\includegraphics[width=\unitlength,page=1]{speaker_amp_output_attack_ratio.pdf}}%
  \end{picture}%
\endgroup%

%% file: Figures/plot/speaker_comparator_output.pdf_tex
\begingroup%
  \makeatletter%
  \providecommand\color[2][]{%
    \errmessage{(Inkscape) Color is used for the text in Inkscape, but the package 'color.sty' is not loaded}%
    \renewcommand\color[2][]{}%
  }%
  \providecommand\transparent[1]{%
    \errmessage{(Inkscape) Transparency is used (non-zero) for the text in Inkscape, but the package 'transparent.sty' is not loaded}%
    \renewcommand\transparent[1]{}%
  }%
  \providecommand\rotatebox[2]{#2}%
  \newcommand*\fsize{\dimexpr\f@size pt\relax}%
  \newcommand*\lineheight[1]{\fontsize{\fsize}{#1\fsize}\selectfont}%
  \ifx\svgwidth\undefined%
    \setlength{\unitlength}{403.54893135bp}%
    \ifx\svgscale\undefined%
      \relax%
    \else%
      \setlength{\unitlength}{\unitlength * \real{\svgscale}}%
    \fi%
  \else%
    \setlength{\unitlength}{\svgwidth}%
  \fi%
  \global\let\svgwidth\undefined%
  \global\let\svgscale\undefined%
  \makeatother%
  \begin{picture}(1,0.70794832)%
    \lineheight{1}%
    \setlength\tabcolsep{0pt}%
    \put(0,0){\includegraphics[width=\unitlength,page=1]{speaker_comparator_output.pdf}}%
  \end{picture}%
\endgroup%

%% file: Figures/block/experiment_setup.pdf_tex
\begingroup%
  \makeatletter%
  \providecommand\color[2][]{%
    \errmessage{(Inkscape) Color is used for the text in Inkscape, but the package 'color.sty' is not loaded}%
    \renewcommand\color[2][]{}%
  }%
  \providecommand\transparent[1]{%
    \errmessage{(Inkscape) Transparency is used (non-zero) for the text in Inkscape, but the package 'transparent.sty' is not loaded}%
    \renewcommand\transparent[1]{}%
  }%
  \providecommand\rotatebox[2]{#2}%
  \newcommand*\fsize{\dimexpr\f@size pt\relax}%
  \newcommand*\lineheight[1]{\fontsize{\fsize}{#1\fsize}\selectfont}%
  \ifx\svgwidth\undefined%
    \setlength{\unitlength}{249.78437885bp}%
    \ifx\svgscale\undefined%
      \relax%
    \else%
      \setlength{\unitlength}{\unitlength * \real{\svgscale}}%
    \fi%
  \else%
    \setlength{\unitlength}{\svgwidth}%
  \fi%
  \global\let\svgwidth\undefined%
  \global\let\svgscale\undefined%
  \makeatother%
  \begin{picture}(1,0.38097905)%
    \lineheight{1}%
    \setlength\tabcolsep{0pt}%
    \put(0,0){\includegraphics[width=\unitlength,page=1]{experiment_setup.pdf}}%
  \end{picture}%
\endgroup%

%% file: Figures/plot/motor_comparator_output.pdf_tex
\begingroup%
  \makeatletter%
  \providecommand\color[2][]{%
    \errmessage{(Inkscape) Color is used for the text in Inkscape, but the package 'color.sty' is not loaded}%
    \renewcommand\color[2][]{}%
  }%
  \providecommand\transparent[1]{%
    \errmessage{(Inkscape) Transparency is used (non-zero) for the text in Inkscape, but the package 'transparent.sty' is not loaded}%
    \renewcommand\transparent[1]{}%
  }%
  \providecommand\rotatebox[2]{#2}%
  \newcommand*\fsize{\dimexpr\f@size pt\relax}%
  \newcommand*\lineheight[1]{\fontsize{\fsize}{#1\fsize}\selectfont}%
  \ifx\svgwidth\undefined%
    \setlength{\unitlength}{405.6918146bp}%
    \ifx\svgscale\undefined%
      \relax%
    \else%
      \setlength{\unitlength}{\unitlength * \real{\svgscale}}%
    \fi%
  \else%
    \setlength{\unitlength}{\svgwidth}%
  \fi%
  \global\let\svgwidth\undefined%
  \global\let\svgscale\undefined%
  \makeatother%
  \begin{picture}(1,0.70684992)%
    \lineheight{1}%
    \setlength\tabcolsep{0pt}%
    \put(0,0){\includegraphics[width=\unitlength,page=1]{motor_comparator_output.pdf}}%
  \end{picture}%
\endgroup%

%% file: Figures/block/timeline.pdf_tex
\begingroup%
  \makeatletter%
  \providecommand\color[2][]{%
    \errmessage{(Inkscape) Color is used for the text in Inkscape, but the package 'color.sty' is not loaded}%
    \renewcommand\color[2][]{}%
  }%
  \providecommand\transparent[1]{%
    \errmessage{(Inkscape) Transparency is used (non-zero) for the text in Inkscape, but the package 'transparent.sty' is not loaded}%
    \renewcommand\transparent[1]{}%
  }%
  \providecommand\rotatebox[2]{#2}%
  \newcommand*\fsize{\dimexpr\f@size pt\relax}%
  \newcommand*\lineheight[1]{\fontsize{\fsize}{#1\fsize}\selectfont}%
  \ifx\svgwidth\undefined%
    \setlength{\unitlength}{351.08337962bp}%
    \ifx\svgscale\undefined%
      \relax%
    \else%
      \setlength{\unitlength}{\unitlength * \real{\svgscale}}%
    \fi%
  \else%
    \setlength{\unitlength}{\svgwidth}%
  \fi%
  \global\let\svgwidth\undefined%
  \global\let\svgscale\undefined%
  \makeatother%
  \begin{picture}(1,0.28238922)%
    \lineheight{1}%
    \setlength\tabcolsep{0pt}%
    \put(0,0){\includegraphics[width=\unitlength,page=1]{timeline.pdf}}%
  \end{picture}%
\endgroup%